\documentclass[journal,comsoc]{IEEEtran}
\usepackage{amsfonts}
\usepackage{amssymb}
\usepackage{eurosym}
\usepackage{cite}
\usepackage{graphicx}
\usepackage{epstopdf}
\usepackage{amsmath}
\usepackage{tikz,lipsum}
\usepackage{caption}
\usepackage[T1]{fontenc}
\usepackage{amsthm}
\usepackage{mathrsfs}
\usepackage{color}
\usepackage{stfloats}
\usepackage{xcolor, etoolbox}
\usepackage{subcaption}

\IEEEoverridecommandlockouts
\newcounter{mytempeqncnt}
\newtheorem{remark}{Remark}

\newtheorem{proposition}{Proposition}

\begin{document}

\title{On the Secrecy Enhancement of an Integrated Ground-Aerial Network
with a Hybrid FSO/THz Feeder Link}
\author{Elmehdi Illi, \IEEEmembership{Member, IEEE}, and Marwa Qaraqe, %
\IEEEmembership{Senior Member, IEEE} \thanks{%
E. Illi and M. Qaraqe are with the College of Science and Engineering, Hamad
Bin Khalifa University, Qatar Foundation, Doha, Qatar. (e-mails:
elmehdi.illi@ieee.org, mqaraqe@hbku.edu.qa.} }
\maketitle

\begin{abstract}
High altitude platforms (HAPs)-aided terrestrial-aerial communication technology based on free-space optical\ (FSO) and Terahertz (THz) feeder links has been attracting notable interest recently due to its great potential in reaching a higher data rate and connectivity. Nonetheless, the presence of harsh vertical propagation environments and potential aerial eavesdroppers are two of the main challenges limiting the reliability and security of such a technology. In this work, a secrecy-enhancing scheme for HAP-aided ground-aerial communication is
proposed. The considered network consists of HAP-assisted communication between a ground station and a legitimate user under the threat of an aerial and ground eavesdropper. Thus, the proposed scheme leverages (i) HAP diversity by exploiting the presence of multiple flying HAPs and (ii) the use of a hybrid FSO/THz transmission scheme to offer better
resilience against eavesdropping attacks. An analytical secrecy outage probability (SOP) expression is derived for the scheme in consideration. Results manifest the notable gain in security of the proposed scheme with respect to both (i) the single-HAP and (ii)\ THz feeder-based benchmark ones, where the proposed scheme's SOP\ is decreased by four orders of
magnitude using $4$ HAPs with respect to the first benchmark scheme, while a $5$-dB secrecy gain is manifested with respect to the second benchmark one.
\end{abstract}
\section{Introduction}

Throughout the past few years, there has been an increasing interest in exploiting the aerial interface, composed of unmanned aircraft, high-altitude platforms (HAPs), and satellites, for extending the area coverage of several underserved zones over the earth and significantly increasing the peak data rate \cite{slimfrontiers}. HAPs and satellites can broaden communication coverage
and boost data rates through multi-beam transmissions and by leveraging very high-throughput backhaul/feeder links from an earth station. In this optic, HAPs manifested several benefits compared to conventional low-earth orbit (LEO) satellites, essentially its shorter signal round-trip time compared to their LEO counterpart due to the shorter distance to the earth
surfaces, i.e., placed in the stratosphere (17-50 Km), hence providing a low communication latency \cite{gunessurvey}.

From another front, the adoption of the THz and FSO transmissions as a backhaul feeder link for terrestrial-aerial/satellite transmissions has been among the advocated solutions to cater to the high data rate needs. Such technologies can solve the spectrum saturation issue in the radio frequency (RF) spectrum in the Ku and Ka bands \cite{gunessurvey,ojcoms}. FSO is based on transmitting conical-shaped optical beams in the infrared (IR)/near IR or visible light spectrum. Its immunity to interference, security, high amount of bandwidth, and low implementation cost make FSO feeder links a viable solution for the continuous needs in bandwidth. On the other hand, THz technology is based on the utilization of the 0.1 THz-10\ THz RF\ spectrum for transmitting directive RF\ beams \cite{akyildiz}. To this end, THz technology can offer at least an order of magnitude of bandwidth gain compared to its mmWave counterpart. However, despite the aforementioned features, the two technologies suffer from several common impairments, such as beam wandering, atmospheric turbulence-induced fading, and, essentially, beam pointing errors due to the transceivers' vibration/movement. In addition to this, it has been shown that FSO suffers heavily from meteorological conditions, such as the presence of fog, heavy rain, and clouds, whereas THz manifests better resilience against severe weather. It is worth mentioning that the presence of molecular absorption is an additional impairment affecting the THz
communication's performance, while the FSO\ transmission is less affected by it.

The physical layer security (PLS) paradigm has been attracting the wireless community in the past years. PLS'\ objective is to ensure secure keyless transmission, from either a confidentiality or authentication point of view, relying solely on the physical layer parameters, e.g., fading, antenna diversity, and precoding. Confidentiality-based PLS relies on maximizing/enhancing the secrecy capacity (SC) metric to counter eavesdropping attacks, which can guarantee a higher transmission rate as well as a target decoding failure at eavesdroppers. From FSO\ and THz communications point of view, in spite of their high beam directivity, exhibiting "by nature"\ a secure transmission, THz and optical beams are of divergent nature at higher transmission distances. Precisely, the beam spot size at an aerial receiver in the stratosphere can reach the order of tens-hundreds of meters, which puts it under a continuous threat of malicious aircraft/unmanned aerial vehicles (UAVs) that can be located within the beam divergence area
and eavesdrop on the ground-space/satellite feeder link \cite%
{martinez,fsosecrecy}. Therefore, securing the feeder link in hybrid terrestrial-aerial/satellite links is crucial.

\subsection{Related Work}


Concerning the literature on FSO-based terrestrial-aerial/satellite networks, several works inspected its performance in terms of bit error and outage probabilities and link margin, as detailed in \cite{ahmad,lit10,lit15}. In addition to this, the authors of \cite{jornet} explored the potential of THz communication in airplane-satellite links, whereby a holistic channel modeling encompassing the various propagation attenuation phenomena and a performance evaluation of the system were provided. On the other hand, several techniques have been proposed to enhance the performance of ground-aerial/satellite networks. For instance, the authors in \cite{lit11} proposed the incorporation of site diversity by using multiple ground stations to counter the high attenuation of ground-satellite FSO links. Furthermore, in \cite{lit12,lit13}, another approach was proposed to enhance the reliability performance of hybrid terrestrial-aerial/satellite networks by adopting a hybrid RF/FSO transmission, where the system switches to the RF\ Ka-band signals whenever the received FSO signal power falls below a threshold limit. Furthermore, in \cite{lit14}, a HAPs diversity scheme was proposed to increase the reliability of downlink satellite-HAP-earth networks by selecting the best relaying HAP in terms of the instantaneous signal-to-noise ratio (SNR).

From another front, some research works inspected the secrecy of FSO- and THz-based terrestrial-aerial/satellite networks. For instance, the secrecy of a dual-hop terrestrial-satellite network was analyzed in \cite{ojcoms} with an FSO feeder link and under the presence of an eavesdropper in each of the two hops, i.e., uplink ground-satellite and downlink satellite-ground. In order to increase the secrecy of the considered network, both optical aperture diversity and RF\ beams precoding were considered. Also, Ma et al. tackled in \cite{lit8} the secrecy evaluation of a mixed RF-FSO uplink terrestrial-satellite connection assisted by a relay UAV. In addition, the authors of \cite{lit1} analyzed the security level of a satellite- and HAP-based network under two communication scenarios, namely the satellite-to-HAP and HAP-to-ground under eavesdropping attacks. The same authors inspected in \cite{lit2} the secrecy of a satellite-ground communication system assisted by a HAP as a relay, where an eavesdropper attempts to compromise the second hop operating with RF\ Ka-band. In \cite{lit3}, the security of an RF-based reconfigurable intelligent surface (RIS)-and UAV-aided satellite-ground communication was analyzed in the presence of multiple ground eavesdroppers. The work of Ben Yahia et al. in \cite{lit5} aimed to inspect the security of an FSO-based satellite-HAP\ network when a malicious spacecraft attempts to overhear the legitimate signal from the divergent optical beam. The same authors elaborated an extension of the aforementioned work in \cite{lit6}, where the analysis of secrecy considered FSO-based satellite-HAP, uplink ground-HAP, and downlink HAP-ground networks in the presence of a HAP eavesdropper. In \cite{lit9}, a cognitive radio-based terrestrial-satellite network with an optical feeder link was quantified in terms of the key system parameters. It is worth highlighting notable existing works that quantified the secrecy level of terrestrial FSO\ communication systems under different eavesdropping situations and positions, such as in \cite{martinez,fsosecrecy}. Lastly, from the THz-based vertical networks' point of view, the work in \cite{lit7} provided a thorough secrecy investigation of a THz-based downlink satellite-ground communication aided by a HAP. A RIS is mounted onboard the HAP\ to act as a passive relay and beamsteer the THz beam towards a ground user, under the presence of an eavesdropper in its vicinity.

\subsection{Motivation}

It has been established from the above-discussed literature work that the
reliability performance of FSO- and THz-based hybrid terrestrial-aerial/satellite networks is heavily impacted by atmospheric attenuation, turbulence, and beam misalignment (pointing errors). The
proposed reliability enhancing techniques, such as HAP/site diversity and
hybrid RF/FSO\ transmission, have been established to fulfill a higher
reliability target based on the received SNR\ as a metric \cite {lit11,lit12,lit13,lit14}. On the other hand, from a network secrecy point of view, most prior secrecy investigations considered eavesdropping threats solely on the RF\ side of the network,
i.e., generally in the fronthaul (radio access) satellite/HAP-to-earth
station (user) links. Also, the work in \cite{lit1,lit5,lit6,lit7} showed that the system's secrecy can exhibit some weakness in several scenarios. Essentially, the FSO, THz, and RF\ links' impairments in such vertical links significantly affect their respective secrecy. Therefore, it is of paramount
importance to design secrecy-enhancing schemes for terrestrial-aerial/satellite networks to fulfill better data reliability (higher SNR\ levels) and also higher resilience against eavesdropping attacks from potential aerial eavesdroppers.

Motivated by the above, the current work aims to propose a novel PLS
scheme for hybrid terrestrial-aerial networks by exploiting both HAPs\
diversity alongside a hybrid FSO/THz-feeder link transmission. The
considered network consists of a HAP-aided communication between a ground station (GS)\ and a ground user, whereby the GS-HAP communication operates mainly over an FSO\ link due to its very high data rates, whereas the HAP decodes and forwards the received signal to the ground user. Under the potential presence of an eavesdropping HAP and a ground eavesdropper aiming to compromise both transmission hops, the proposed scheme aims at the exploitation of multiple available flying HAPs within the reach of the legitimate transmit ground station (GS) to select the one maximizing its secrecy performance under the presence of several aerial eavesdroppers and a single terrestrial one. In addition to this, the proposed scheme relies on a hybrid FSO/THz GS-HAPs feeder link, whereby THz transmission is activated as a backup link whenever the FSO\ transmission fails in fulfilling a target secrecy level. The use of THz as a backup link is expected to enhance the secrecy performance of the system further, in addition to HAP diversity, due to its resilience to propagation phenomena compared to its FSO counterpart. The current work exhibits the main differences with respect to the aforementioned previous works, where in \cite{lit11,lit12,lit13,lit14}, the performance of the considered networks was analyzed from reliability's point of view and not analyzing the secrecy considering either HAPs diversity \cite{lit14} or hybrid RF/FSO links \cite{lit12,lit13}, while in \cite{lit1,lit2,lit3,lit4,lit5,lit6,lit7,lit8,ojcoms}, the secrecy performance was inspected by considering FSO-feeder links and a single relay/HAP/satellite. To the best of our knowledge, the current work is the first to propose the use of HAPs diversity and hybrid FSO/THz transmission for the secrecy enhancement of HAP-aided terrestrial-aerial networks.

\subsection{Contributions}

The main contributions of the current work can be summarized as follows:

\begin{itemize}
\item A novel secrecy-enhancing scheme is proposed by harnessing the
benefits of HAPs diversity and hybrid THz/FSO\ transmission.

\item A\ novel mathematical framework for evaluating the network's secrecy
outage probability (SOP)\ metric is derived, encompassing key system
parameters.

\item Several analytical observations are provided to demonstrate
analytically the effect of some system parameters.

\item Extensive numerical simulations are conducted to analyze the effect of key system parameters on the system's secrecy performance.
\end{itemize}

The rest of this paper is organized as follows:\ Section II presents the
considered system and channel model, while Section III provides useful
statistical functions for the secrecy evaluation. In Section\ IV, the
proposed scheme is detailed, and its secrecy performance is derived. Section V\ is dedicated to showing illustrative numerical results for the system's secrecy. Lastly, Section VI concludes the paper.

\section{System Model}

\begin{figure}[h]
\vspace*{-.1cm}

\par
\begin{center}
\includegraphics[scale=.31]{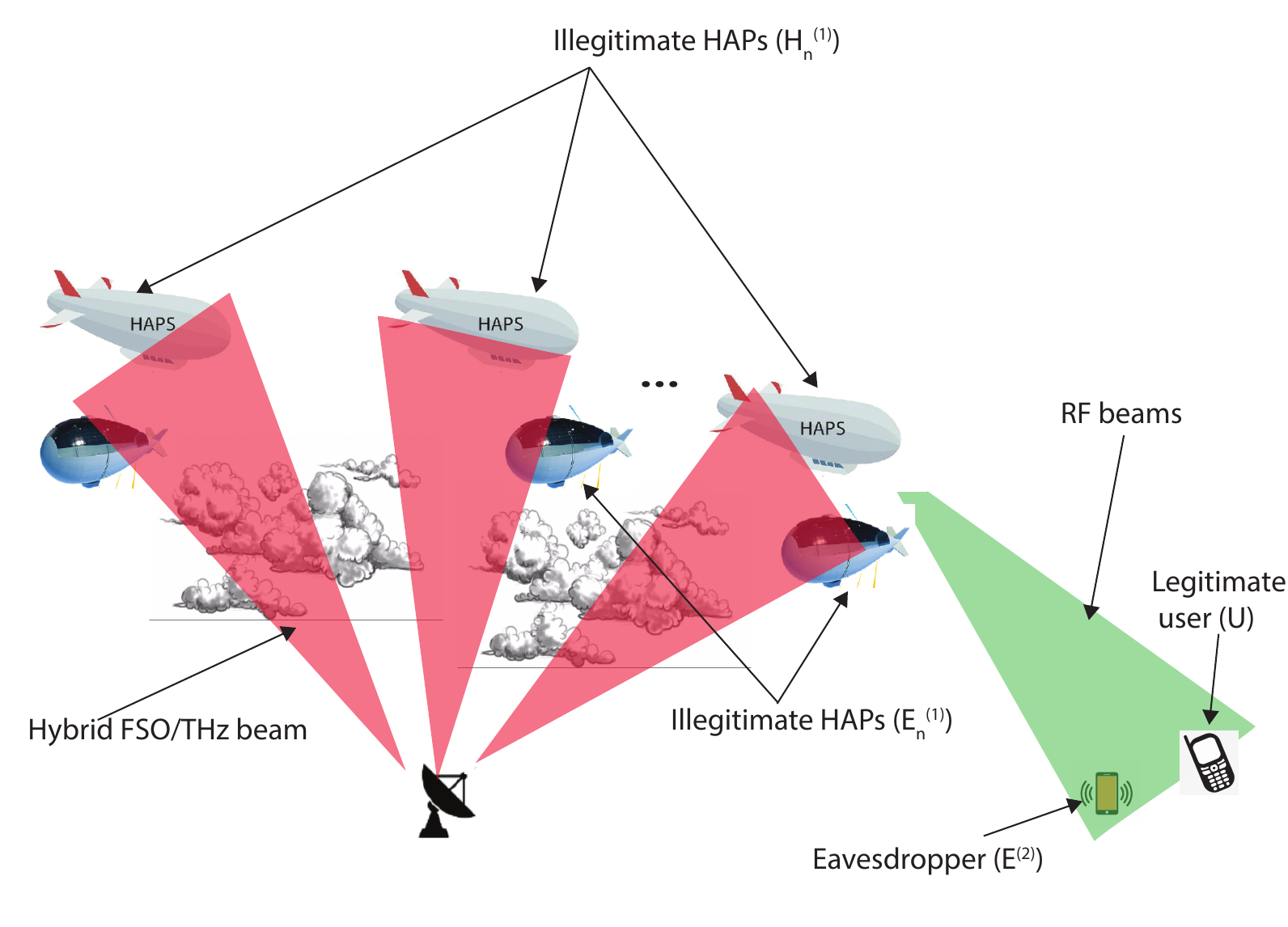}
\end{center}
\par
\caption{System model}
\label{sysmod}
\end{figure}

	The transmission from the GS\ at the earth to an end-user is carried out through the help of $N$ HAPs $\left\{ H_{n}\right\} _{n=1,\ldots ,N}$\
acting as relays, as shown in Fig. \ref{sysmod}. FSO beams convey the
transmit data symbols of the ground user over a turbulent uplink vertical
channel impaired by atmospheric turbulence, scattering caused by
particles, free-space path loss, and the potential presence of clouds, rain,
or fog. In addition to this, the beam misalignment between the ground
station and each relay HAP causes inevitable pointing errors. Furthermore, multiple eavesdropping HAPs $\left\{ E_{n}^{(1)}\right\} _{n=1,\ldots ,N}$ are targeting the legitimate information beams, where each malicious HAP is located at the vicinity of a legitimate one, precisely in the optical beam divergence area, to decode the legitimate signal. Due to the presence of the aforementioned impairments, the FSO\ link might become unreliable at some transmission slots and manifests a very high power loss. To this end, a backup THz link is adopted to ensure a backhaul communication between the GS and the HAP\ in the case of the FSO\ link's severe attenuation. The GS\ selects a HAP\ among the $N$\ HAPs according to the highest secrecy level, which will be detailed in the next section. Afterward, the selected HAP\ decodes the received information signal over either the THz or FSO\ link, regenerates it, and forwards it to a ground user $U$ using an RF\ link operating over the Ka-band. It is assumed that the second hop's communication is under the threat of a malicious ground eavesdropper $E^{(2)} $ that targets the illegitimate interception of the RF\ beam.

\subsubsection{FSO\ Link}

The optical wireless transmission between the earth and the flying HAPs is established using either intensity modulation with direct detection (IM/DD) or coherent heterodyne detection (CHD) one. While the former is established by modulating the information using the light intensity of the optical signal, the latter technique conveys the information symbols using
the amplitude (i.e., intensity), phase, or frequency. In addition to this,
each eavesdropper is assumed to capture a portion $\rho _{E}$ of the optical
beam's power that was not captured by the corresponding legitimate HAP, as
proposed in \cite{martinez}. Without loss of generality, it is assumed that
the optical power is shared between the legitimate and illegitimate HAPs
such that $\rho _{H_{n}}=1-\rho _{E_{n}}$, with $\rho _{H_{n}}$ indicating
the portion of power captured by the $n$th legitimate HAP $H_{n}$. To this
end, the received optical signal at the $n$th legitimate/illegitimate HAP,
using both aforementioned techniques can be expressed as follows \cite%
{ojcoms}%
\begin{equation}
y_{X_{n}}^{\text{(FSO)}}=\left( \eta \rho _{X}I_{SX_{n}}\right) ^{\frac{r}{2}%
}x+w_{X_{n}},X\in \left\{ H,E^{(1)}\right\}
\end{equation}%
where $\eta $ is the electrical-to-optical conversion efficiency in A/W, $%
I_{SX_{n}}$ is the received light irradiance, $x$ is the transmit signal
with an average optical power $E\left[ x\right] =P_{o}$, $r\in \left\{
1,2\right\} $ is a detection-dependent parameter that equals $1$ for CHD and
$2$ for the IM/DD, and $w_{X_{n}}$ is the additive white Gaussian noise
process (AWGN) at the receiving HAP of zero mean and variance $\sigma
_{F_{n}}^{2}$. The received light irradiance generally results from
three main signal attenuation phenomena, namely:

\begin{enumerate}
\item The atmospheric attenuation, denoted by $I_{SX_{n}}^{(l)}$, due to the
scattering by air particles throughout the propagation path, rain/fog
attenuation, and free-space path-loss.

\item The induced fading caused by atmospheric turbulence, i.e, $%
I_{SX_{n}}^{(a)}$.

\item The pointing errors fluctuating attenuation caused by transceivers
misalignment, i.e., $I_{SX_{n}}^{(p)}$.
\end{enumerate}

Consequently, the resulting light irradiance expression can be formulated as
follows: $
I_{SX_{n}}=I_{SX_{n}}^{(l)}I_{SX_{n}}^{(a)}I_{SX_{n}}^{(p)}$. 
The path-loss term is mainly caused by scattering loss through the small
droplets in rain, fog, and clouds, and other particles in the atmosphere.
In addition to this, propagation's free-space path loss attenuates the
received optical signal power. As a result, the aggregate optical
attenuation can be expressed as\ \cite{lit14,carbonneu,itufso,itufso2}%
\begin{align}
I_{SX_{n}}^{(l)}& =\exp \left( -\sigma _{n}L_{SX_{n}}^{\text{(eff,cloud/fog
[km])}}\right) \exp \left( -\frac{\tau _{X_{n}}}{\cos \left( \psi
_{SX_{n}}\right) }\right)  \notag \\
& \times 10^{-\delta _{SX_{n}}^{\text{(rain,FSO)}}L_{SX_{n}}^{\text{%
(eff,rain)}}}G_{S}^{\text{(T,FSO)}}G_{X_{n}}^{\text{(R,FSO)}}\left( \frac{%
\lambda _{\text{FSO}}}{4\pi L_{SX_{n}}}\right) ^{2},  \label{pathlosstot}
\end{align}%
where $\sigma $ is the geometric scattering extinction coefficient due to
cloud or fog droplets, which can be evaluated as follows \cite{lit9}: $\sigma _{n}=\left( \frac{3.91}{V_{n}}\right) \left( \frac{\lambda _{\text{FSO%
}}^{\text{[nm]}}}{550}\right) ^{-q}$,
where $V_{n}=\frac{1.002}{\left( KQ\right) ^{0.6473}} $
is the visibility in kilometers, $K$ is the liquid water content in g/m$%
^{-3} $ and $Q$ indicates the cloud contentration in m$^{-3}$.\ Furthermore,
$\lambda _{\text{FSO}}^{\text{[nm]}}$ \footnote{%
The notation [x] in the superscript of some length parameters refers to the
parameter value in x units, e.g., $\lambda _{\text{FSO}}^{\text{[nm]}}$
refers to the FSO signal wavelength in nanometers. Not specifying a unit in
the superscript indicates the use of meters as a unit.} is the FSO\ signal's
wavelength in micrometers, and $q$ is the particle size-related coefficient
defined per the Kim's model as \cite{fsoweather}%
\begin{equation}
q=\left\{
\begin{array}{l}
1.6,\text{ }V_{n}>50 \\
1.3,\text{ }6<V_{n}\leq 50 \\
0.16V_{n}+0.34,\text{ }1<V_{n}\leq 6 \\
V_{n}-0.5,0.5<V_{n}\leq 1 \\
0,V_{n}\leq 0.5%
\end{array}%
\right.
\end{equation}%
where the visibility bounds in the last equation are in kilometers. Also, $%
L_{X_{n}}^{\text{(eff,cloud/fog)}}$ is the effective scattering length
between the GS and the $n$th benign/malign HAP spanning only along the
actual cloud or fog thickness, defined as follows \cite{alouinimag}%
\begin{equation}
L_{X_{n}}^{\text{(eff,cloud/fog, [km])}}=\frac{\Delta L^{\text{(cloud/fog,
[km])}}}{\cos \left( \psi _{SX_{n}}\right) }
\end{equation}%
where $\Delta L^{\text{(cloud/fog, [km])}}$ is the cloud/fog layer thickness
and $\psi _{SX_{n}}$ is the zenith angle between the GS and the $n$th
legitimate/illegitimate HAP. On the other hand, $\xi _{X_{n}}$ refers to the
Mie scattering extinction coefficient caused by microscopic water particles
at the sea surface, and is expressed as follows
\begin{align}
\tau _{X_{n}}& =p_{1}\left( \lambda _{\text{FSO}}\right) \left[ h_{S}^{\text{%
[km]}}\right] ^{3}+p_{2}\left( \lambda _{\text{FSO}}\right) \left[ h_{S}^{%
\text{[km]}}\right] ^{2}  \notag \\
& +p_{3}\left( \lambda _{\text{FSO}}\right) h_{S}^{\text{[km]}}+p_{4}\left(
\lambda _{\text{FSO}}\right)
\end{align}%
where $h_{S}^{\text{[km]}}$ is the GS altitude above the sea level in
kilometers, and $p_{i}\left( \lambda \right) $ $(i=1,\ldots ,4)$ are
wavelength-dependent parameters defined as \cite{itufso}%
\begin{align}
p_{1}\left( \lambda _{\text{FSO}}\right) & =0.000487\left( \lambda _{\text{%
FSO}}^{\text{[}\mu \text{m]}}\right) ^{3}-0.002237\left( \lambda _{\text{FSO}%
}^{\text{[}\mu \text{m]}}\right) ^{2}  \notag \\
& +0.003864\lambda _{\text{FSO}}^{\text{[}\mu \text{m]}}-0.004442,
\end{align}%
\begin{align}
p_{2}\left( \lambda _{\text{FSO}}\right) & =-0.00573\left( \lambda _{\text{%
FSO}}^{\text{[}\mu \text{m]}}\right) ^{3}+0.02639\left( \lambda _{\text{FSO}%
}^{\text{[}\mu \text{m]}}\right) ^{2}  \notag \\
& -0.04552\lambda _{\text{FSO}}^{\text{[}\mu \text{m]}}+0.05164,
\end{align}%
\begin{eqnarray}
p_{3}\left( \lambda _{\text{FSO}}\right) &=&0.02565\left( \lambda _{\text{FSO%
}}^{\text{[}\mu \text{m]}}\right) ^{3}-0.1191\left( \lambda _{\text{FSO}}^{%
\text{[}\mu \text{m]}}\right) ^{2}  \notag \\
&&+0.20385\left( \lambda _{\text{FSO}}^{\text{[}\mu \text{m]}}\right) -0.216,
\end{eqnarray}%
\begin{eqnarray}
p_{4}\left( \lambda _{\text{FSO}}\right) &=&-0.0638\left( \lambda _{\text{FSO%
}}^{\text{[}\mu \text{m]}}\right) ^{3}+0.3034\left( \lambda _{\text{FSO}}^{%
\text{[}\mu \text{m]}}\right) ^{2}  \notag \\
&&-0.5083\lambda _{\text{FSO}}^{\text{[}\mu \text{m]}}+0.425.
\end{eqnarray}%
Rain attenuation contributes also to the overall path loss in vertical FSO\
links. In practice, rain droplets' size increases in heavy rain situations
which causes additional photon refractions and scattering \cite{lit9}.
Carbonneau deployed an analytical model by leveraging on empirical
observations of rain attenuation in FSO\ transmissions in \cite{carbonneu}.
The underlying rain attenuation coefficient is represented by $\delta
_{SX_{n}}^{\text{(rain,FSO)}}=1.076\mathcal{R}^{0.67}$ (dB/km), where $%
\mathcal{R}$ is the rain rate in mm/hr. It is worth highlighting that,
equivalently to the cloud/fog attenuation, the rain attenuation is
considered only along the rain layer thickness, as given in (\ref%
{pathlosstot}) where $L_{SX_{n}}^{\text{(eff,rain,[km])}}$ is the optical
signal propagation length throughout the rain layer, defined as \cite%
{alouinimag}%
\begin{equation}
L_{SX_{n}}^{\text{(eff,rain,[km])}}=\frac{\Delta L^{\text{(rain,[km])}}}{%
\cos \left( \psi _{SX_{n}}\right) },
\end{equation}%
where $\Delta L^{\text{(rain,[km])}}$ is the rain layer thickness in km. On
top of that, $G^{\text{(T,THz)}}$and $G_{X_{n}}^{\text{(R,THz)}}$ denote the
optical gains for the GS' transmit laser and the receiving HAP's
photodetector, respectively, and $L_{X_{n}}$ is the GS-$n$th
legitimate/illegitimate HAP link distance. Lastly, the

As far as the turbulence-induced fading (i.e., $I_{SX_{n}}^{(a)}$) is
concerned, Gamma-Gamma distribution has been widely adopted as a unifying
distribution manifesting a wide range of turbulence regimes (i.e., from weak
to strong turbulence). On the other hand, the Rayleigh pointing error model
has shown a notable accuracy in representing the statistics of the fluctuation
due to pointing error provoked by beam misalignment $\left( \text{i.e., }%
I_{SX_{n}}^{(p)}\right) $.\ The next section details the mathematical
representation of the received irradiance and SNR's statistics.

The receiver performs either direct detection to decode the information
symbol relying on the light intensity or the coherent heterodyne one whereby
either the optical signal's intensity (i.e., amplitude), phase, or frequency
modulates the information symbol. To this end, the received signal-to-noise
ratio (SNR) at the $n$th legitimate/illegitimate HAP is given as

\begin{equation}
\gamma _{1,X_{n}}^{\text{(FSO)}}=\frac{\left( \eta \rho
_{X_{n}}I_{SX_{n}}\right) ^{r}P_{F}}{\sigma _{X_{n},\text{F}}^{2}},
\label{snrfso}
\end{equation}%
where $P_{F}$ is the average transmit electrical power of the GS\ over the
FSO link.

\subsection{\protect THz Link}

When the ground-to-HAP\ transmission operates with the THz band, the
propagation is affected essentially by molecular absorption, atmospheric
attenuation due to cloud/fog and rain, free-space path loss, and pointing
errors. As a result, the received SNR\ at the $n$th legitimate/illegitimate
HAP\ can be expressed as

\begin{equation}
\gamma _{1,X_{n}}^{\text{(THz)}}=\frac{P_{T}\mathcal{L}_{SX_{n}}\mathcal{%
\delta }_{SX_{n}}^{\text{(rain,THz)}}\mathcal{\delta }_{SX_{n}}^{\text{%
(cloud/fog,THz)}}}{\exp \left( \kappa _{a}\left( f\right) L_{SX_{n}}\right)
\left\vert h_{SX_{n}}\right\vert ^{2}\sigma _{X_{n},\text{T}}^{2}},
\label{snrthz}
\end{equation}%
where $P_{T}$ is the average transmit THz signal power,%
\begin{equation}
\mathcal{L}_{SX_{n}}=G^{\text{(T,THz)}}G_{X_{n}}^{\text{(R,THz)}}\left(
\frac{\lambda _{\text{THz}}}{4\pi L_{SX_{n}}}\right) ^{2},
\end{equation}%
is the free-space path loss term,
\begin{equation}
\mathcal{\delta }_{SX_{n}}^{\text{(rain,THz)}}\triangleq 10^{-\frac{%
\vartheta \left( \lambda _{\text{THz}}\right) \mathcal{R}^{\nu (\lambda _{%
\text{THz}})}L_{X_{n}}^{\text{(eff, rain, [km])}}}{10}}
\end{equation}%
is the rain attenuation, where $\mathcal{R}$ is the attenuation rate, $%
\vartheta \left( \lambda _{\text{THz}}\right) $ and $\nu \left( \lambda _{%
\text{THz}}\right) $ are medium- and wavelength-dependent constants for the
rain attenuation,
\begin{equation}
\mathcal{\delta }_{SX_{n}}^{\text{(cloud)}}\triangleq 10^{-KM\left( \lambda
_{\text{THz}}\right) L_{X_{n}}^{\text{(eff, cloud/fog, [km])}}}
\end{equation}%
is the cloud/fog attenuation coefficient, where $M\left( \lambda _{\text{THz}%
}\right) $ is medium- and wavelength-dependent cloud attenuation
coefficient. The values for the parameters $\vartheta \left( .\right) $ and $%
\nu \left( .\right) $ can be computed from \cite[Eqs. (4), (5), Table 5]%
{iturain} for the considered THz frequency, while the cloud/fog attenuation
parameter $M\left( \lambda _{\text{THz}}\right) $ can be computed using{{{\
\cite[Eqs. (2)-(11)]{itucloud}. }}}In addition, $\kappa _{a}\left( f\right) $
is the frequency-dependent absorption coefficient in km$^{-1}$, and $%
h_{SX_{n}}$ is the pointing errors random fluctuations term.

\subsection{Second Hop:\ mmWave Link}

After decoding the received signal and regenerating it, the selected
legitimate HAP forwards it to the legitimate ground user $U$, under the
presence of a malicious eavesdropping node attempting to overhear the
legitimate RF\ signal beam. The propagation over the RF\ Ka-Band is less
affected by molecular absorption, and beams are usually of large footprints,
allowing eavesdroppers to intercept the legitimate message with higher
probability. Therefore, RF\ transmission is not impacted by pointing errors.
Nonetheless, it is worth highlighting that multipath fading and shadowing
affect RF\ vertical links. To this end, the SNR at the
legitimate/illegitimate ground receiver, given the signal is received from
the $n$th HAP, is formulated as

\begin{equation}
\gamma _{2,Z}^{\text{(Ka)}}=\frac{P_{H}\mathcal{\delta }_{H_{n}Z}^{\text{%
(rain,Ka)}}\mathcal{\delta }_{H_{n}Z}^{\text{(cloud/fog,Ka)}}\mathcal{L}%
_{H_{n}Z}\left\vert h_{H_{n}Z}\right\vert ^{2}}{\sigma _{Z}^{2}},Z\in
\{U,E^{(2)}\},  \label{snrka}
\end{equation}%
\newline
where $P_{H}$ is the HAP\ transmit power, $\mathcal{\delta }_{H_{n}Z}^{\text{%
(rain,Ka)}}$ and $\mathcal{\delta }_{H_{n}Z}^{\text{(cloud/fog,Ka)}}$ are,
Similar to the THz link, the respective rain and cloud/fog attenuation of
the RF link. Also,

\begin{equation}
\mathcal{L}_{H_{n}Z}=G_{H_{n}}^{\text{(T,Ka)}}G_{Z}^{\text{(R,Ka)}}\left(
\frac{\lambda _{\text{Ka}}}{4\pi L_{H_{n}Z}}\right) ^{2}
\end{equation}%
is the path-loss of the HAP-user link, $h_{H_{n}Z}$ is the complex-valued
shadowed fading coefficient, modeled by the shadowed Rician distribution,
and $\sigma _{Z}^{2}$ is the additive white Gaussian noise at the receiver.

\section{Useful Statistics}

In this section, statistical properties of the received SNR\ of each
particular link is detailed, which will be used in the next section for the
system's secrecy evaluation.

\subsection{\protect FSO\ Link}

The PDF\ and cumulative distribution
function (CDF) of the SNR $\gamma _{1,X_{n}}^{\text{(FSO)}}$ in (\ref{snrfso}) can be expressed as \cite[Eqs. (14)-(15)]{ojcoms}%
\begin{align}
f_{\gamma _{1,X_{n}}^{\text{(FSO)}}}(z)& =\frac{\xi _{X_{n},\text{FSO}}^{2}}{%
rz\Gamma \left( \alpha _{X_{n}}\right) \Gamma \left( \beta _{X_{n}}\right) }
\notag \\
& \times G_{1,3}^{3,0}\left( \left( \frac{\mathcal{F}_{X_{n},1}z}{\overline{%
\gamma }_{X_{n}}^{\text{(FSO)}}}\right) ^{\frac{1}{r}}\left\vert
\begin{array}{c}
-;\xi _{X_{n},\text{FSO}}^{2}+1 \\
\xi _{X_{n},\text{FSO}}^{2},\alpha _{X_{n}},\beta _{X_{n}};-%
\end{array}%
\right. \right) ,  \label{pdffso}
\end{align}%
and%
\begin{align}
F_{\gamma _{1,X_{n}}^{\text{(FSO)}}}(z)& =\frac{r^{\alpha _{X_{n}}+\beta
_{X_{n}}-2}\xi _{X_{n},\text{FSO}}^{2}}{\left( 2\pi \right) ^{r-1}\Gamma
\left( \alpha _{X_{n}}\right) \Gamma \left( \beta _{X,n}\right) }  \notag \\
& \times G_{r+1,3r+1}^{3r,1}\left( \frac{\mathcal{F}_{X_{n},r}z}{\overline{%
\gamma }_{X_{n}}^{\text{(FSO)}}}\left\vert
\begin{array}{c}
1;\varepsilon _{1}^{(X_{n})} \\
\varepsilon _{2}^{(X_{n})};0%
\end{array}%
\right. \right) ,  \label{cdffso}
\end{align}%
respectively,
for $X\in \left\{ H,E^{(1)}\right\} $, where
\begin{equation}
\xi _{X_{n},\text{FSO}}\triangleq \frac{w_{z_{eq}}^{\text{(FSO)}}}{2\sigma
_{s}},  \label{xi}
\end{equation}%
accounts for the severity of the pointing errors, where%
\begin{equation}
w_{z_{eq}}^{\text{(FSO)}}=w_{z}^{\text{(FSO)}}\sqrt{\frac{\sqrt{\frac{\pi }{%
A_{0}^{\text{(FSO)}}}}}{2v\exp \left( -v_{\text{FSO}}^{2}\right) }}
\label{wzeq}
\end{equation}%
refers to equivalent beam radius at the receiver plane {{\cite{farid}}}
where $w_{z}^{\text{(FSO)}}$ is the FSO\ beam waist at the receiver's plane,
\begin{equation}
A_{0}^{\text{(FSO)}}\triangleq \mathrm{erf}^{2}\left( v_{\text{FSO}}\right)
\label{A0}
\end{equation}%
is the fraction of collected power in pointing error-free transmission,
\begin{equation}
v_{\text{FSO}}\triangleq \sqrt{\pi /2}R_{\text{FSO}}/w_{z}^{\text{(FSO)}},
\label{v}
\end{equation}%
$R_{\text{FSO}}$ is the HAP's photodetector radius, and $\sigma _{s}$ is the
receiver's jitter standard deviation along the $x$ and $y$ axes.
Additionally, {{$\alpha _{X_{n}}$ and $\beta _{X_{n}}$ are the
turbulence-induced fading parameters, defined for the case of untracked
uplink beam as {\cite[Eq. ]{andrews}}}}%
\begin{equation}
{{\alpha _{X_{n}}=\left[
\begin{array}{c}
5.95\left( \frac{2w_{0}^{\text{(FSO)}}}{r_{0}}\right) ^{\frac{5}{3}}\left(
\frac{\sigma _{pe}}{w_{z_{eq}}^{\text{(FSO)}}}\right) ^{2} \\
+\exp \left( \frac{0.49{{\sigma _{R,X_{n}}^{2}}}}{\left( 1+0.56{{\sigma
_{R,X_{n}}^{12/5}}}\right) ^{7/6}}\right) -1%
\end{array}%
\right] ^{-1}}}
\end{equation}%
and%
\begin{equation*}
{{\beta _{X,n}=}}\left[ \exp \left( \frac{0.51{{\sigma _{R,X_{n}}^{2}}}}{%
\left( 1+0.69{{\sigma _{R,X_{n}}^{12/5}}}\right) ^{5/6}}\right) -1\right]
^{-1},
\end{equation*}%
respectively, {where }$w_{0}$ is the beam waist at the transmit GS,
\begin{equation}
r_{0}=\left[ 0.42k^{2}\sec \left( \psi _{X_{n}}\right)
\int_{h_{G}}^{h_{X_{n}}}C_{n}^{2}\left( h\right) dh\right] ^{-3/5}
\end{equation}%
is the atmospheric coherence diameter with $k=2\pi /\lambda _{\text{FSO}}$
being the wave number, and%
\begin{align}
C_{n}^{2}\left( h\right) & =0.00594\left( \mathcal{V}/27\right) ^{2}\left(
10^{-5}h\right) ^{10}\exp \left( -\frac{h}{1000}\right)  \notag \\
& +2.7\times 10^{-16}\exp \left( -\frac{h}{1500}\right) +C_{0}\exp \left(
-h/100\right) ,
\end{align}%
is the refractive index structure parameter in m$^{-2/3}$, where $\mathcal{V}
$ is root mean square value of the wind speed, and $C_{0}$ is the refractive
index structure parameter value at the GS' altitude above sea level.
Additionally,
\begin{equation}
\sigma _{pe}=0.54\left( \frac{L_{SX_{n}}\lambda _{\text{FSO}}}{2w_{0}^{%
\text{(FSO)}}}\right) ^{2} \frac{  1-\left( \frac{\left( \frac{2\pi w_{0}^{\text{(FSO)}}}{r_{0}}%
\right) ^{2}}{\left( \frac{2\pi w_{0}^{\text{(FSO)}}}{r_{0}}\right) ^{2}+1}%
\right) ^{1/6} }{\left( \frac{2w_{0}^{\text{(FSO)}}}{r_{0}}\right)
^{\frac{-5}{3}}} .
\end{equation}%
Also,%
\begin{align}
{{\sigma _{R,X_{n}}^{2}}}& {=2.25k}{}^{\frac{7}{6}}\sec ^{\frac{11}{6}%
}\left( \psi _{SX_{n}}\right) \int_{h_{S}}^{h_{X_{n}}}\left( 1-\frac{h-h_{S}%
}{h_{X_{n}}-h_{S}}\right) ^{\frac{5}{6}}  \notag \\
& \times C_{n}^{2}\left( h\right) \left( h-h_{S}\right) ^{\frac{5}{6}}dh{,}
\end{align}%
is the Rytov variance of the propagation system, representing the turbulence
severity level, with $h_{S}$ and $h_{{X}_{n}}$ representing the altitudes of
the GS and $X_{n}$ above the sea level, respectively. Furthermore,\ $%
G_{p,q}^{m,n}\left( .\left\vert .\right. \right) $ is the Meijer's $G$%
-function \cite[Eq. (07.34.02.0001.01)]{wolfram}, $
\mathcal{F}_{X_{n},r}\triangleq \left( \frac{\xi _{X_{n},\text{FSO}%
}^{2}\alpha _{X_{n}}\beta _{X_{n}}}{r^{2}\left( \xi _{X_{n},\text{FSO}%
}^{2}+1\right) }\right) ^{r}$,  $\varepsilon _{1}^{(X_{n})}\triangleq \left( \frac{\xi
_{X_{n},\text{FSO}}^{2}+1}{r}\right) _{i=1,..,r}$, $\varepsilon
_{2}^{(X_{n})}\triangleq \left( \frac{\xi _{X_{n},\text{FSO}}^{2}+i}{r},%
\frac{\alpha _{X_{n}}+i}{r},\frac{\beta _{X,n}+i}{r}\right) _{i=0,..,r-1}$,
and
\begin{equation}
\overline{\gamma }_{X_{n}}^{\text{(FSO)}}=\frac{P_{F}\left( \eta \rho
_{X_{n}}I_{SX_{n}}^{(l)}\right) ^{r}\mathbb{E}^{r}\left[ I_{SX_{n}}^{(ap)}%
\right] }{\sigma _{X_{n},\text{F}}^{2}}  \label{avgsnrfso}
\end{equation}%
is the corresponding average electrical SNR with {{\cite{ansarimalaga}}}
\begin{equation}
\mathbb{E}\left[ I_{SX_{n}}^{(ap)}\right] =\frac{\xi _{X_{n},\text{FSO}%
}^{2}A_{0}^{\text{(FSO)}}}{\xi _{X_{n},\text{FSO}}^{2}+1}.
\label{mompointingfso}
\end{equation}

\subsection{THz Link}

Similarly to the FSO\ link, the THz channel is impacted by
Rayleigh-distributed pointing errors, whereby the corresponding attenuation
attenuation's magnitude, i.e., $\left\vert h_{SX_{n}}\right\vert $, has the
following PDF
\begin{equation}
f_{\left\vert h_{SX_{n}}\right\vert }\left( x\right) =\frac{\xi _{X_{n},%
\text{THz}}^{2}}{\left( A_{0}^{\text{(THz)}}\right) ^{\xi _{X_{n},\text{THz}%
}^{2}}}x^{\xi _{X_{n},\text{THz}}^{2}-1},0<x<A_{0}^{\text{(THz)}},
\label{pdfthzfading}
\end{equation}%
where $\xi _{X_{n},\text{THz}}^{2}$ is the pointing errors severity for the
THz link, which can be computed equivalently to the FSO\ channel using (\ref%
{xi})-(\ref{v}), by substituting the FSO beam waist by the THz beam's one,
and the photodetector's size by the THz antenna's physical aperture. As a
consequence, armed by the Jacobian transform as well as some algebraic
manipulations, one can retrieve the PDF and CDF of the THz link's SNR,
expressed in (\ref{snrthz}), as%
\begin{equation}
f_{\gamma _{1,X_{n}}^{\text{(THz)}}}\left( x\right) =\left\{
\begin{array}{l}
\frac{\xi _{X_{n},\text{THz}}^{2}\left( \frac{x}{\overline{\gamma }%
_{1,X_{n}}^{\text{(THz)}}}\right) ^{\frac{\xi _{X_{n},\text{THz}}^{2}}{2}-1}%
}{2\left( A_{0}^{\text{(THz)}}\right) ^{\xi _{X_{n},\text{THz}}^{2}}%
\overline{\gamma }_{1,X_{n}}^{\text{(THz)}}},\text{ if }0<x\leq \gamma
_{1,X_{n}}^{\text{(THz,max)}} \\
0,\text{elsewhere}%
\end{array}%
\right. ,  \label{pdfthz}
\end{equation}%
and%
\begin{equation}
F_{\gamma _{1,X_{n}}^{\text{(THz)}}}\left( x\right) =\left\{
\begin{array}{l}
\frac{\left( \frac{x}{\overline{\gamma }_{1,X_{n}}^{\text{(THz)}}}\right) ^{%
\frac{\xi _{X_{n},\text{THz}}^{2}}{2}}}{\left( A_{0}^{\text{(THz)}}\right)
^{\xi _{X_{n},\text{THz}}^{2}}},\text{ if }0<x\leq \gamma _{1,X_{n}}^{\text{%
(THz,max)}} \\
1,\text{elsewhere}%
\end{array}%
\right. ,  \label{cdfthz}
\end{equation}%
respectively, where
$
\gamma _{1,X_{n}}^{\text{(THz,max)}}\triangleq \overline{\gamma }_{1,X_{n}}^{%
\text{(THz)}}\left( A_{0}^{\text{(THz)}}\right) ^{2}$
and%
\begin{equation}
\overline{\gamma }_{1,X_{n}}^{\text{(THz)}}\triangleq \frac{P_{T}\mathcal{L}%
_{SX_{n}}\mathcal{\delta }_{SX_{n}}^{\text{(rain,THz)}}\mathcal{\delta }%
_{SX_{n}}^{\text{(cloud/fog,THz)}}\mathbb{E}\left[ \left\vert
h_{SX_{n}}\right\vert ^{2}\right] }{\exp \left( \kappa _{a}\left( f\right)
L_{SX_{n}}\right) \sigma _{X_{n},\text{T}}^{2}}  \label{avgsnrthz}
\end{equation}%
is the average THz link's SNR with
\begin{equation}
\mathbb{E}\left[ \left\vert h_{SX_{n}}\right\vert ^{2}\right] =\frac{\xi
_{X_{n},\text{THz}}^{2}\left( A_{0}^{\text{(THz)}}\right) ^{2}}{\xi _{X_{n},%
\text{THz}}^{2}+2}.  \label{mompointingthz}
\end{equation}

\begin{remark}
\label{pointingeffect} It can be noted from equations (\ref{avgsnrfso}), (%
\ref{mompointingfso}), (\ref{avgsnrthz}), and (\ref{mompointingthz}) that
the average fading power due to the pointing errors for both the FSO\ and
THz links depends on the value $\xi _{X_{n},\text{x}}$ $\left( \text{x}\in
\left\{ \text{THz,\ FSO}\right\} \right) $, defined in (\ref{xi}). This
latter is proportional to the equivalent beam waist radius $w_{z_{eq}}^{%
\text{(x)}}$, which is per se an proportional to the beam waist (spot size)\
at the receiver, as shown in (\ref{wzeq}). Therefore, the higher the higher
the received beam's spot size at the receiver (higher $\xi _{X_{n},\text{x}}$
), the greater is the corresponding moments in (\ref{mompointingfso}), (\ref%
{mompointingthz}) and average SNR levels in (\ref{avgsnrfso}), (\ref%
{avgsnrthz}). Consequently, this results in an increased average received
SNR value.
\end{remark}

\subsection{MmWave Link}

 The HAP-to-ground user communication is performed using Ka-band RF\
signals that are subject to shadowed fading impairments. Abdi et al.
developed in \cite{abdi} a theoretical model to represent the distribution
of signal attenuation in a land mobile-satellite link, for which the
underlying PDF and CDF of the SNR in (\ref{snrka})\ is given as \cite[Eqs.
(18)-(19)]{ojcoms}%
\begin{align}
f_{\gamma _{2,Z}}\left( x\right) & =\frac{\phi _{Z}}{\overline{\gamma }_{2,Z}%
}\exp \left( -\frac{v_{Z}x}{\overline{\gamma }_{2,Z}}\right)  \sum\limits_{n=0}^{m_{s}^{(Z)}-1}\frac{\binom{m_{s}^{(Z)}-1}{n} \left( \frac{\mu _{Z}x}{\overline{\gamma }_{2,Z}}\right) ^{n}}{n!%
}  \label{pdfka}
\end{align}%
and%
\begin{equation}
F_{\gamma _{2,Z}}\left( z\right) =\phi
_{Z}\sum\limits_{n=0}^{m_{s}^{(Z)}-1}\binom{m_{s}^{(Z)}-1}{n}\frac{\mu
_{Z}^{n}\gamma _{\text{inc}}\left( n+1,\frac{\mu _{Z}x}{\overline{\gamma }%
_{2,Z}}\right)}{v_{Z}^{n+1}n!}    ,  \label{cdfka}
\end{equation}
respectively, where $Z\in \left\{
U,E^{(2)}\right\}$, $\phi _{Z}{\triangleq }${{$\frac{1}{2b}\left(
\frac{2bm_{s}^{(Z)}}{2b_{Z}m_{s}^{(Z)}+\Omega _{s}^{(Z)}}\right) ^{m_{s}}$,}}%
\begin{equation}
\overline{\gamma }_{2,Z}=\frac{P_{H}\mathcal{\delta }_{H_{n}Z}^{\text{%
(rain,Ka)}}\mathcal{\delta }_{H_{n}Z}^{\text{(cloud/fog,Ka)}}\mathcal{L}%
_{H_{n}Z}\mathbb{E}\left[ \left\vert h_{H_{n}Z}\right\vert ^{2}\right] }{%
\sigma _{Z}^{2}},
\end{equation}%
is the average received SNR\ at the legitimate/illegitimate receiver, {{$%
v_{Z}\triangleq {\zeta }_{Z}-\mu _{Z}$, }}${\zeta }${{$_{Z}\triangleq \frac{1%
}{2b_{Z}}$,}} {{$2b_{Z}$ is the average power of multipath components, and }}%
$\mu ${{$_{Z}{\triangleq }\frac{\Omega _{s}^{(Z)}}{2b_{Z}\left(
2b_{Z}m_{s}^{(Z)}+\Omega _{s}^{(Z)}\right) }$. Additionally, }}$\Omega
_{s}^{(Z)}${\ is the average power of the LOS\ components, }$m_{s}^{(Z)}$\
is the fading severity parameter, and{\ }$\gamma _{\text{inc}}\left(
.,.\right) ${{\ is the lower-incomplete Gamma function \cite[Eq. (9.210)]%
{integrals}}}.{\ }

\section{Proposed Scheme and Secrecy Analysis}

In this section, the proposed HAP selection scheme and the respective
secrecy evaluation is detailed. In the considered network, the ground
station (GS) aims at selecting one relaying HAP among the $N$ available
ones, along with the suitable transmission link (i.e., either the FSO or
THz)\ that exhibits the highest SC. It is assumed that prior to each
transmission instant, the GS\ has knowledge of the instantaneous channel
state information of the $S$-$H_{n}$, $S$-$E_{n}^{(1)}$, $H_{n}$-$U$, and $%
H_{n}$-$E^{(2)}$ links, which can be used to compute the respective links' SNR\ levels.

\subsection{\protect Communication Secrecy: Background}

From PLS' perspective, a confidential transmission exists if and only if the
legitimate channel's capacity exceeds the malicious one's. Such a capacity
difference defines the well-known SC\ metric as%
\begin{equation}
C_{s}=\left[ C_{L}-C_{E}\right] ^{+},  \label{csdef}
\end{equation}%
where the indices $L$ and $E$ refer to any pair of legitimate and
illegitimate transceivers/entities in a wireless network, where%
\begin{equation}
C_{L/E}=\log _{2}\left( 1+\gamma _{L/E}\right)  \label{capdef}
\end{equation}%
and $\gamma _{L/E}$ denote the malign/benign link's bandwidth-normalized
channel capacity and $\gamma _{L/E}$, respectively, and $\left[ x\right]
^{+}=\max \left( 0,x\right) $. The SC\ defines the maximal transmission rate
that can guarantee both communication reliability, i.e., successful message
decoding at the legitimate receiver, and a unit equivocation rate at the
eavesdropper \cite{wyner}. It can be noted from (\ref{csdef})\ that the
greater the legitimate link's capacity (i.e., SNR) and/or the lower the
malign ones, the better the SC, thus allowing for higher secure transmission rates.

The SC is randomly fluctuating according to the wireless channel (i.e., SNR)
statistics. Thus, in practice, a communication rate is fixed such that the
SC\ unlikely falls below it. If the latter scenario occurs, a secrecy outage
event takes place, where the preset communication rate $R_{s}$ cannot fail
in ensuring either data reliability at the legitimate receiver or the target
equivocation at the eavesdropper. Therefore, the SOP\ is defined as \cite%
{sustainable}%
\begin{eqnarray}
P_{s} &\triangleq &\Pr \left[ C_{s}<R_{s}\right]  \notag \\
&=&\Pr \left[ \gamma _{L}<2^{R_{s}}\left( 1+\gamma _{E}\right) -1\right] .
\label{sopdef}
\end{eqnarray}%
Such a probability can be expressed in terms of the PDF\ and CDF\ of the
illegitimate channel's SNR\ and the legitimate's one, respectively, as%
\begin{equation}
P_{s}=\int_{\mathcal{D}_{\gamma _{E}}}F_{\gamma _{L}}\left( 2^{R_{s}}\left(
1+z\right) -1\right) f_{\gamma _{E}}\left( z\right) dz.
\label{sopdefintegral}
\end{equation}

The interval $\mathcal{D}_{\gamma _{E}}$ defines the range of the
eavesdropper's SNR\ $\gamma _{E}$. Typically, $\mathcal{D}_{\gamma _{E}}=%
\mathbb{R}
^{+}$ for most of propagation scenarios. Nonetheless, it should be pointed
out that the SNR\ can exhibit some limit values, such as THz communication
subject to pointing errors, whereby the SNR exhibits the limit $\overline{%
\gamma }_{1,X_{n}}^{\text{(THz)}}\left( A_{0}^{\text{(THz)}}\right) ^{2}$.

\subsection{Proposed Scheme}

The proposed scheme's objective is to select a single HAP\ out of the $N$
available ones, along with the transmission link (i.e., either THz or FSO),
exhibiting the best SC. Without loss of generality, it is assumed that a
perfect CSI\ knowledge of all the channels is available at GS for HAP/link
selection. The proposed scheme's process can be detailed as follows:

\begin{enumerate}
\item The GS inspects the CSI of the channels with respect to the $N$
legitimate and illegitimate HAPs over both the THz and FSO\ links, which can
be obtained by feedback links. At the same time, each HAP estimates its CSI\
from the legitimate and illegitimate ground receivers through the same
mechanism.

\item Then, the measured second-hop CSI\ values are sent back to the GS to
compute the corresponding legitimate and illegitimate SNRs, and,
consequently, the secrecy capacity of each link using (\ref{csdef}) and (\ref%
{capdef}).

\item Then, for each of the $N$ haps, the GS\ selects the link $\ell \in \{$%
THz, FSO$\}$ according to the following rule%
\begin{equation}
\ell =\left\{
\begin{array}{l}
\text{FSO, if \ }C_{s,1,\text{FSO}}^{(n)}\geq R_{s} \\
\text{THz, if \ }C_{s,1,\text{FSO}}^{(n)}<R_{s},C_{s,1,\text{THz}}^{(n)}\geq
R_{s}%
\end{array}%
\right.
\end{equation}%
where $C_{s,1,\text{FSO}}^{(n)}$ and $C_{s,1,\text{THz}}^{(n)}$ are the
respective FSO\ and THz channels' SCs, which can be expressed using (\ref%
{csdef}), (\ref{capdef}), and the SNR\ expression in (\ref{snrfso}) and (\ref%
{snrthz}). It is worth mentioning that the third case when both SCs fall
below $R_{s}$ is identified as a communication outage, where transmission is
suspended.

\item The GS\ selects the best HAP $H_{n^{\ast }}$ encompassing the highest
end-to-end (e2e)\ SC as%
\begin{equation}
n^{\ast }=\arg \max_{n=1,\ldots ,N}C_{s}^{(n)},
\end{equation}%
where
\begin{equation}
C_{s}^{(n)}=\min \left( C_{s,1,\ell }^{(n)},C_{s,2,\text{Ka}}^{(n)}\right)
\end{equation}%
is the e2e SC\ of the dual-hop link via the $n$th HAP operating with
decode-and-forward\ (DF) relaying and $C_{s,2,\text{Ka}}^{(n)}$ is the
corresponding second hop's SC over the RF\ link.
\end{enumerate}

\subsection{\protect Secrecy Evaluation of the Proposed Scheme}

\begin{proposition}
The SOP\ of the proposed scheme can be formulated as
\begin{equation}
P_{s}=\prod\limits_{n=1}^{N}\left[ P_{s,\text{FSO}}^{(1,n)}P_{s,\text{THz}%
}^{(1,n)}+P_{s,\text{Ka}}^{(2,n)}-P_{s,\text{FSO}}^{(1,n)}P_{s,\text{THz}%
}^{(1,n)}P_{s,\text{Ka}}^{(2,n)}\right] .  \label{soptotal}
\end{equation}%
where $P_{s,\text{FSO}}^{(1,n)}$, $P_{s,\text{THz}}^{(1,n)}$, and $P_{s,%
\text{Ka}}^{(2,n)}$ indicate the SOP\ of the FSO, THz, and Ka-band RF\ links

\begin{proof}
Kindly refer to Appendix A.
\end{proof}
\end{proposition}

\begin{remark}
It can be seen from equation (\ref{soptotal}) that the overall system's SOP\
is expressed in terms of the individual per-link and per-hop SOP,
independently. Therefore, the quantification of each of the underlying
link's SOP\ can provide an overall system's SOP\ evaluation.
\end{remark}

\begin{proposition}
The SOP\ of the individual FSO, THz, and RF links, given the communication
is performed through the $n$th HAP, can be expressed in (\ref{sopfsofinal}),
(\ref{sopthzfinal}), and (\ref{soprffinal}), respectively, shown at the top
of the current and next page
\begin{figure*}[t]
{\normalsize 
\setcounter{mytempeqncnt}{\value{equation}}
}
\par
\begin{align}
P_{s,\text{FSO}}^{(1,n)}& =\frac{r^{\alpha _{E_{n}}+\beta _{E_{n}}+\alpha
_{H_{n}}+\beta _{H_{n}}-3}\xi _{E_{n},\text{FSO}}^{2}\xi _{H_{n},\text{FSO}%
}^{2}}{r\left( 2\pi \right) ^{2\left( r-1\right) }\Gamma \left( \alpha
_{H_{n}}\right) \Gamma \left( \beta _{H_{n}}\right) \Gamma \left( \alpha
_{E_{n}}\right) \Gamma \left( \beta _{E_{n}}\right) }  \notag \\
& \times \left[
\begin{array}{c}
G_{0,1;r+2,3r+1;r+1,3r}^{1,0;3r,1;3r,1}\left( Q_{H_{n}},Q_{E_{n}}\left\vert
\begin{array}{c}
-;-:1;\varepsilon _{1}^{(H_{n})},0:1;\varepsilon _{1}^{(E_{n})} \\
1;-:\varepsilon _{2}^{(H_{n})};0:\varepsilon _{2}^{(E_{n})};-%
\end{array}%
\right. \right) \\
-\frac{\prod\limits_{i=0}^{r-1}\Gamma \left( \frac{\xi _{E_{n},\text{FSO}%
}^{2}+i}{r}\right) \Gamma \left( \frac{\alpha _{E_{n}}+i}{r}\right) \Gamma
\left( \frac{\beta _{E_{n}}+i}{r}\right) }{\prod\limits_{i=1}^{r}\Gamma
\left( \frac{\xi _{E_{n},\text{FSO}}^{2}+i}{r}\right) }G_{3r+1;3r+1}^{3r,1}%
\left( Q_{H_{n}},Q_{E_{n}}\left\vert
\begin{array}{c}
1;\varepsilon _{1}^{(H_{n})} \\
\varepsilon _{2}^{(H_{n})};0%
\end{array}%
\right. \right)%
\end{array}%
\right]  \label{sopfsofinal}
\end{align}%
\par
{\normalsize 
\hrulefill 
\vspace*{4pt} }
\end{figure*}
\begin{figure*}[t]
{\normalsize 
\setcounter{mytempeqncnt}{\value{equation}}
}
\par
\begin{equation}
P_{s,\text{THz}}^{(1,n)}=\left\{
\begin{array}{l}
\frac{\xi _{E,n,\text{THz}}^{2}\left( \overline{\gamma }_{1,H_{n}}^{\text{%
(THz)}}\right) ^{-\frac{\xi _{H,n,\text{THz}}^{2}}{2}}}{2\left( A_{0}^{\text{%
(THz)}}\right) ^{\xi _{H,n,\text{THz}}^{2}+\xi _{E,n,\text{THz}}^{2}}\left(
\overline{\gamma }_{1,E_{n}}^{\text{(THz)}}\right) ^{\frac{\xi _{E_{n},n,%
\text{THz}}^{2}}{2}}} \\
\times \text{ }_{2}F_{1}\left( -\frac{\xi _{E_{n},n,\text{THz}}^{2}}{2},%
\frac{\xi _{E_{n},n,\text{THz}}^{2}}{2};\frac{\xi _{E_{n},n,\text{THz}}^{2}}{%
2}+1;-\frac{2^{R_{s}}\min \left( \Psi _{1,H_{n}},\gamma _{1,E_{n}}^{\text{%
(THz,max)}}\right) }{2^{R_{s}}-1}\right) \\
+F_{\gamma _{SE_{n}}^{\text{(THz)}}}\left( \gamma _{1,E_{n}}^{\text{(THz,max)%
}}\right) -F_{\gamma _{SE_{n}}^{\text{(THz)}}}\left( \Psi _{1,H_{n}}\right)
\text{, if }\Psi _{1,H_{n}}>0 \\
1,\text{ if }\Psi _{1,H_{n}}<0\text{ }%
\end{array}%
\right. \text{ }  \label{sopthzfinal}
\end{equation}%
\par
{\normalsize 
\hrulefill 
\vspace*{4pt} }
\end{figure*}
\begin{figure*}[t]
{\normalsize 
\setcounter{mytempeqncnt}{\value{equation}}
}
\par
\begin{eqnarray}
P_{s,\text{Ka}}^{(2,n)} &=&\frac{\phi _{U}\phi _{E^{(2)}}\exp \left( -\frac{%
v_{E^{(2)}}\left[ \frac{1}{2^{R_{s}}}-1\right] }{\overline{\gamma }%
_{2,E^{(2)}}}\right) }{2^{R_{s}}\overline{\gamma }_{2,E^{(2)}}}%
\sum\limits_{n=0}^{m_{s}^{(E^{(2)})}-1}\sum\limits_{p=0}^{m_{s}^{(U)}-1}%
\frac{\binom{m_{s}^{(E^{(2)})}-1}{n}}{n!}\frac{\left( \frac{\mu _{E^{(2)}}}{%
2^{R_{s}}\overline{\gamma }_{2,E^{(2)}}}\right) ^{n}\binom{m_{s}^{(U)}-1}{p}%
\mu _{U}^{p}}{v_{U}^{p+1}}  \notag \\
&&\times \sum\limits_{q=0}^{n}\frac{\binom{n}{q}}{\left( 1-2^{R_{s}}\right)
^{q-n}}\left[
\begin{array}{c}
\left( \frac{\overline{\gamma }_{2,E^{(2)}}2^{R_{s}}}{v_{E^{(2)}}}\right)
^{q+1}\Gamma _{\text{inc}}\left( q+1,\frac{v_{E^{(2)}}\left( 2^{R}-1\right)
}{\overline{\gamma }_{2,E^{(2)}}2^{R_{s}}}\right) \\
-\sum_{k=0}^{p}\frac{\left( \frac{v_{U}}{\overline{\gamma }_{2,U}}\right)
^{k}\Gamma _{\text{inc}}\left( k+q+1,\left( 2^{R_{s}}-1\right) \left[ \frac{%
v_{U}}{\overline{\gamma }_{2,U}}+\frac{v_{E^{(2)}}}{\overline{\gamma }%
_{2,E^{(2)}}2^{R_{s}}}\right] \right) }{\left[ \frac{v_{U}}{\overline{\gamma
}_{2,U}}+\frac{v_{E^{(2)}}}{\overline{\gamma }_{2,E^{(2)}}2^{R}}\right]
^{k+q+1}k!}%
\end{array}%
\right]  \label{soprffinal}
\end{eqnarray}%
\par
{\normalsize 
\hrulefill 
\vspace*{4pt} }
\end{figure*}
where $G_{.,.;.,.;.,.}^{.,.;.,.;.,.}\left( .,.\left\vert .\right. \right) $
is the bivariate Meijer's $G$-function \cite{yakub},
\begin{equation}
Q_{H_{n}}\triangleq \left( \frac{\xi _{H_{n},\text{FSO}}^{2}\alpha
_{H_{n}}\beta _{H_{n}}}{r^{2}\left( \xi _{H_{n},\text{FSO}}^{2}+1\right) }%
\right) ^{r}\frac{2^{R}-1}{\overline{\gamma }_{H_{n}}^{\text{(FSO)}}},
\end{equation}%
\begin{equation}
Q_{E_{n}}\triangleq \left( \frac{\xi _{E_{n},\text{FSO}}^{2}\alpha
_{E_{n}}\beta _{E_{n}}}{r^{2}\left( \xi _{E_{n},\text{FSO}}^{2}+1\right) }%
\right) ^{r}\frac{1-\frac{1}{2^{R}}}{\overline{\gamma }_{E_{n}}^{\text{(FSO)}%
}},
\end{equation}
$_{2}F_{1}\left( .,.;.;.\right) $ is the Gauss hypergeometric function \cite[%
Eq. (07.23.02.0001.01)]{wolfram}, $\Psi _{1,H_{n}}\triangleq \frac{\gamma
_{1,H_{n}^{(1)}}^{\text{(THz,max)}}+1}{2^{R_{s}}}-1$, and $\Gamma _{\text{inc%
}}\left( .,.\right) $ stands for the upper incomplete Gamma function \cite[%
Eq. (8.350.2)]{integrals}.

\begin{proof}
Kindly refer to Appendix B.
\end{proof}
\end{proposition}

\section{Numerical Evaluation}

In this section, numerical results of the proposed scheme's secrecy are
presented to manifest its secrecy performance with respect to the various
system parameters incorporated. Unless otherwise mentioned, the considered
system parameter values are highlighted in Table \ref{paramvals}.
Throughout the results' figures, the analytical curves were plotted by
evaluating the analytical SOP expression by (\ref{soptotal}), (\ref%
{sopfsofinal}), (\ref{sopthzfinal}), and (\ref{soprffinal}). In addition, it
is worth highlighting that Monte Carlo simulation results were obtained by
generating $3\times 10^{6}$ random values to mimic each link's random fading
attenuation, per the distributions in (\ref{pdffso}), (\ref{pdfthz}), and (%
\ref{pdfka}). Also, the numerical results were evaluated by setting $\Upsilon_{H_n}=\Upsilon_{U_n}=\Upsilon$, where $\Upsilon_{H_n}=\frac{P_{\text{F/T}}}{\sigma_{H_n,\text{F/T}}^2}$ is the transmit SNR of the first hop over either the FSO/THz links, while $\Upsilon_{U}=P_H/\sigma_U^2$, whereas $\Upsilon_{E^{(1)}_n}=\Upsilon_{E^{(2)}}=130$ dB.

\begin{table}[!tbp]
\centering
\begin{tabular}{|c|c||c|c|}
\hline\hline
\textbf{Parameter [unit]} & \textbf{Value} & \textbf{Parameter [unit]} &
\textbf{Value} \\ \hline\hline
$\eta $ & $1$ & $G^{\text{(T,FSO)}},G^{\text{(R,FSO)}}$ [dBi] & $10^{12}$
\cite{itufso2} \\ \hline
$\rho _{H_{n}}$ & $2/3$ & $G^{\text{(T,THz)}},G^{\text{(R,THz)}}$ [dBi] & $%
10^{5}$ \\ \hline
$r$ & $2$ & $\lambda _{\text{THz}}$ [mm] & $1.5$ \\ \hline
$\lambda _{\text{FSO}}$ [nm] & $1550$ & $w_{0}^{\text{(FSO)}}$ [cm] & $2$ \\
\hline
$K$ [g/m$^{3}$] & $0.064$ \cite{lit14} & $\mathcal{V}$ \ [m/s] & $21$ \\
\hline
$Q$ [cm$^{-3}$] & $0.025$ \cite{lit14} & $w_{z}^{\text{(FSO)}},w_{z}^{\text{%
(THz)}}$ [m] & $15$ \\ \hline
$\Delta L^{\text{(cloud)}}$ [km] & $5$ & $\sigma _{s}$ [m] & $10$ \\ \hline
$\Delta L^{\text{(fog)}}$ [km] & $0.3$ & $m_{s}^{(Z)}$ & $19$ \\ \hline
$\psi _{SX_{n}},\psi _{H_{n},Z}$ [deg] & $30$ & $\Omega _{s}^{(Z)}$ & $1.29$
\\ \hline
$h_{S}^{\text{[km]}}$ & $0.01$ & $b_{Z}$ & $0.158$ \\ \hline
$\kappa _{a}$ [km$^{-1}$] \cite{ituabsorp} & $4.4\times 10^{-3}$ & $R_{s}$
[bps/Hz] & $3$ \\ \hline
\end{tabular}%
\caption{Simulation parameters' values.}
\label{paramvals}
\end{table}

\begin{figure}[tbp]
\vspace*{-.2cm}
\par
\begin{center}
\includegraphics[scale=0.53]{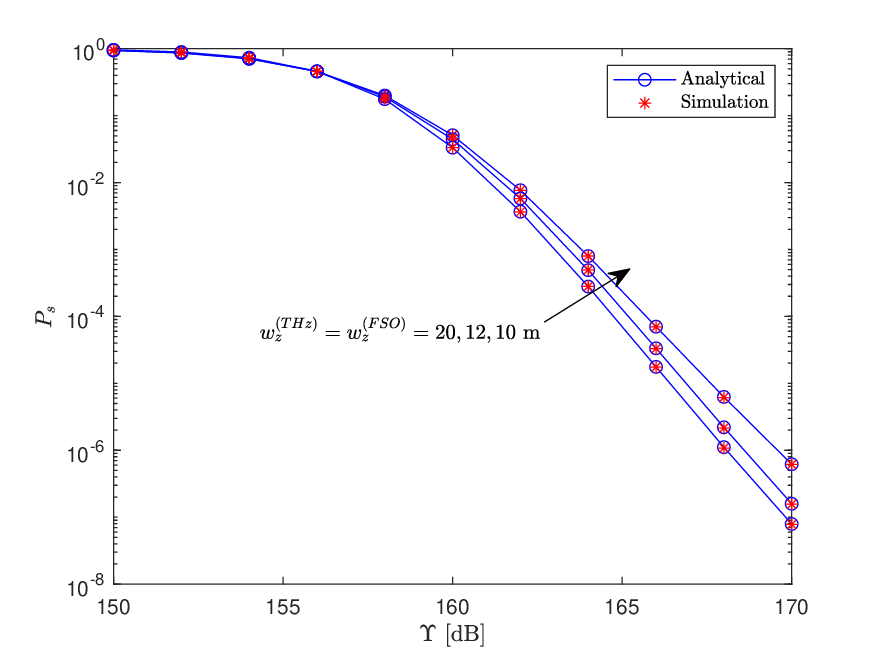}
\end{center}
\par
\vspace*{-.35cm}
\caption{SOP\ of the proposed scheme vs. $\Upsilon $ for different beam
waist values.}
\label{figpointing}
\end{figure}

Fig. \ref{figpointing} presents the system's secrecy in terms of the average
legitimate transmit SNR $\Upsilon $ for different values of the received
beam waist for the FSO and THz links, i.e., $w_{z}^{\text{(FSO)}}$ and $%
w_{z}^{\text{(THz)}}$. It is observed that the system's secrecy is
enhanced i.e., decreasing SOP, with the increase of the beam's spot size at
the receiver. This is due to the fact that the greater the beam waist at the
receiver plane, the smaller the impact of the severity of the pointing errors,
i.e., higher $\xi _{X_{n},\text{FSO}}^{2}$ and $\xi _{X_{n},\text{THz}}^{2}$%
, which results in a higher average received SNR, as shown in Remark \ref%
{pointingeffect}.
\begin{figure}[tbp]
\vspace*{-.2cm}
\par
\begin{center}
\includegraphics[scale=0.53]{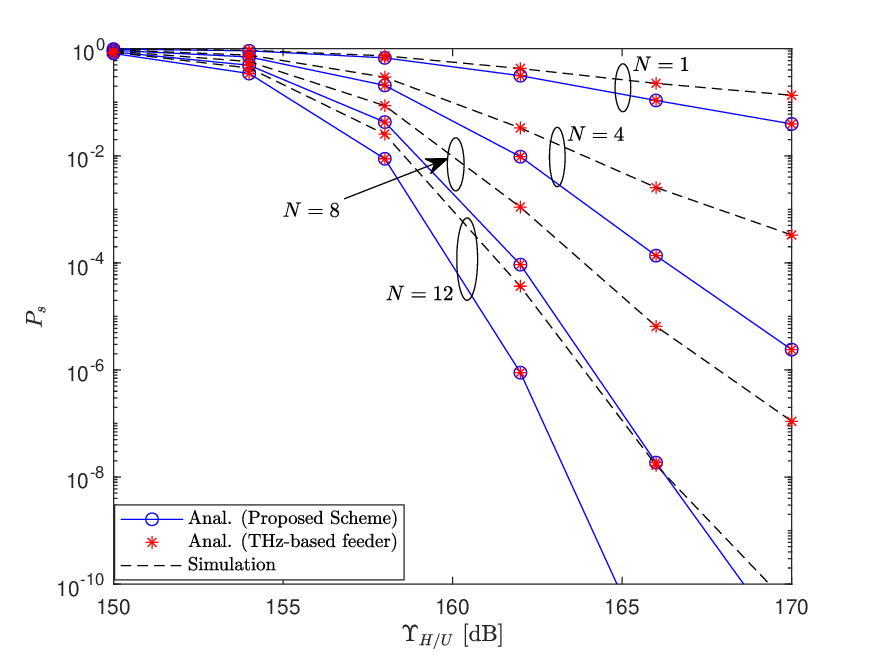}
\end{center}
\par
\vspace*{-.35cm}
\caption{SOP\ of the proposed scheme vs. $\Upsilon $ for different $N$
values.}
\label{figN}
\end{figure}

In Fig. \ref{figN}, the system's SOP performance is shown versus the
legitimate transmit SNR\ $\Upsilon $ for different values of the number of
HAPs $N$. One can ascertain that the increase in the number of HAPs results
in a system's secrecy flourishing, where the SOP\ can reach $10^{-10}$ with $%
\Upsilon =165$ dB and $N=12$. In addition, it can also be noted that at high
$\Upsilon $ values, the SOP\ improves by four orders of magnitude when
increasing the number of HAPs from $N=1$ (benchmark single-HAP system) to $N=4$%
. Furthermore, to gain better insights into the system's secrecy gain, a
comparison with a benchmark scheme relying solely on the THz feeder link is
performed. It is obvious that the proposed hybrid FSO/THz results in better
secrecy enhancement with respect to the THz-feeder-link transmission system,
where the secrecy gain can reach around $4$ and $5$ dB at $P_{s}=10^{-6}$
and $P_{s}=10^{-10}$, respectively.
\begin{figure}[tbp]
\vspace*{-.2cm}
\par
\begin{center}
\includegraphics[scale=0.53]{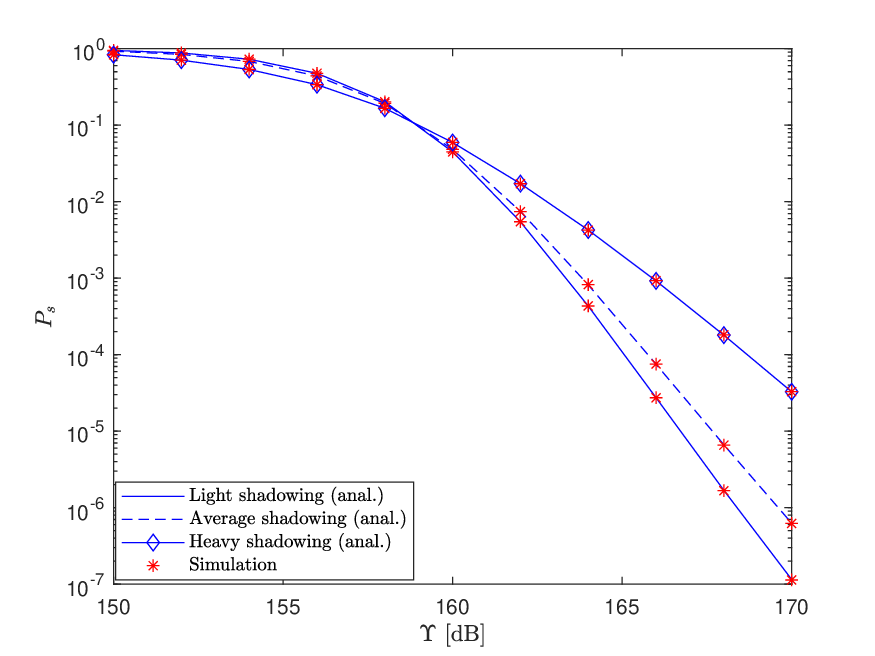}
\end{center}
\par
\vspace*{-.35cm}
\caption{SOP\ of the proposed scheme vs. $\Upsilon $ for different RF\
shadowing scenarios.}
\label{figshadowing}
\end{figure}

The scheme's secrecy performance is shown in Fig. \ref{figshadowing} with
respect to different second hop's RF\ shadowing scenarios, namely light,
average (mild), and strong shadowing scenarios. To this end, the
corresponding fading parameters for the aforementioned scenarios are set,
respectively, as $\left( m_{s}^{(Z)}=19,\Omega
_{s}^{(Z)}=1.29,b_{Z}=0.158\right) $, $\left( m_{s}^{(Z)}=10,\Omega
_{s}^{(Z)}=0.835,b_{Z}=0.126\right) $, and $\left( m_{s}^{(Z)}=1,\Omega
_{s}^{(Z)}=8.97\times 10^{-4},b_{Z}=0.063\right) $, as considered in \cite%
{abdi}. The system's SOP shows an enhancement in the light shadowing regime
where it can reach a level of $10^{-7}$ with $\Upsilon =170$ dB, whereas a
drop by around one and two orders of magnitude is manifested when the second
hop's RF\ propagation is under moderate and strong shadowing cases,
respectively.
\begin{figure}[tbp]
\vspace*{-.2cm}
\par
\begin{center}
\includegraphics[scale=0.53]{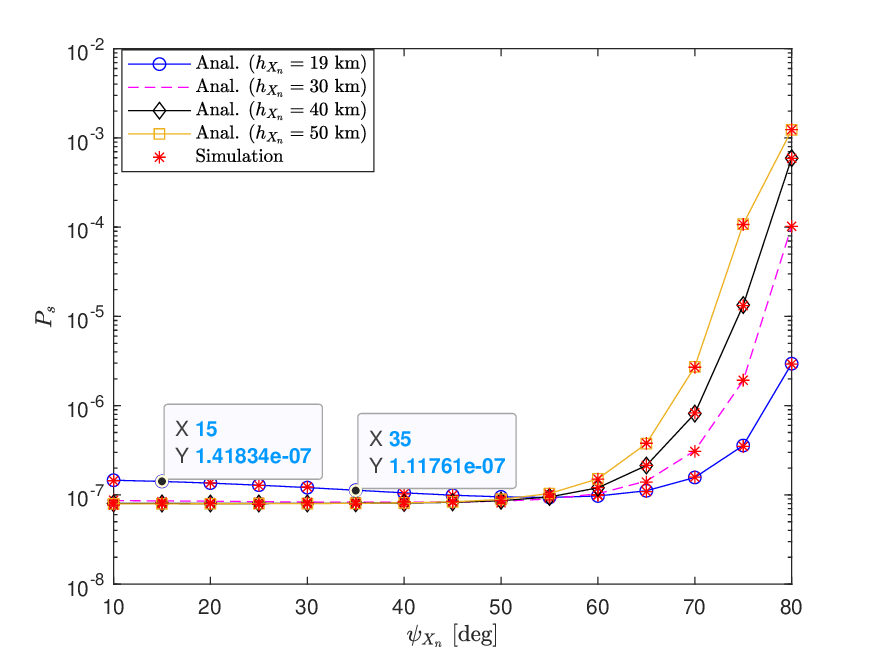}
\end{center}
\par
\vspace*{-.35cm}
\caption{SOP\ of the proposed scheme vs. $\protect\psi _{SX_{n}}$ for
different HAPs altitude.}
\label{figzenith}
\end{figure}

Fig. \ref{figshadowing} presents the scheme's SOP\ in terms of varying
values of the zenith angle ($\psi _{SX_{n}}$) between the GS and the $N$
HAPs. We consider that $\Upsilon =150$ dB, while $\Upsilon
_{E^{(1)}}=\Upsilon _{E^{(2)}}=130$ dB. In order to get more insights on
other physical parameters, such as the HAPs altitude, the performance is
shown for different values of the latter parameter. Several observations can
be made from this figure. For instance, it is ascertained that, for a HAP
altitude of $19$ km, the system's secrecy flourishes in spite of the
increase in $\psi _{SX_{n}}$. This is due to the fact that, for a relatively
lower HAPs altitude, the FSO, THz, and RF\ signals are less affected by the
distance-dependant free-space path loss, which results in a high average
SNR\ levels at both legitimate and illegitimate terminals. The increase in $%
\psi _{SX_{n}}$ between $0^{\circ }$ and $60^{\circ }$ results in a higher
path loss attenuation as the GS-HAP\ distance can be expressed as $%
L_{SX_{n}}=\frac{h_{X_{n}}-h_{S}}{\cos \left( \psi _{SX_{n}}\right) }$.
Therefore, this results in an SNR\ degradation for the legitimate and
illegitimate channels, where one can conclude that the secrecy enhancement
in such a high SNR regime is mainly dominated by degrading the illegitimate
channel's SNR. For $\psi _{SX_{n}}>60^{\circ }$, the SOP\ increases due to
increased path-loss, significantly degrading the SNRs of both types of link.
For instance, a SOP\ of $10^{-3}$ can be reached with $N=4$ HAPs at $50$ km
of altitude. It can be also noted that low HAPs altitude (i.e., high
received average SNR) for lower zenith angle values results in a slight
secrecy degradation compared to higher HAPs altitudes. This is also due to
the fact that for lower zenith angles, a lower altitude yields a high
average SNR\ at the malign and benign HAPs, where the secrecy enhancement is
mainly dominated by the degradation in the illegitimate channel's SNR
(increasing HAPs altitude) rather than the increase of the legitimate
channel's one.

\begin{figure}[tbp]
\centering
\begin{subfigure}[b]{0.5\textwidth}
     \label{case11}
         \centering
         \includegraphics[scale=0.53]{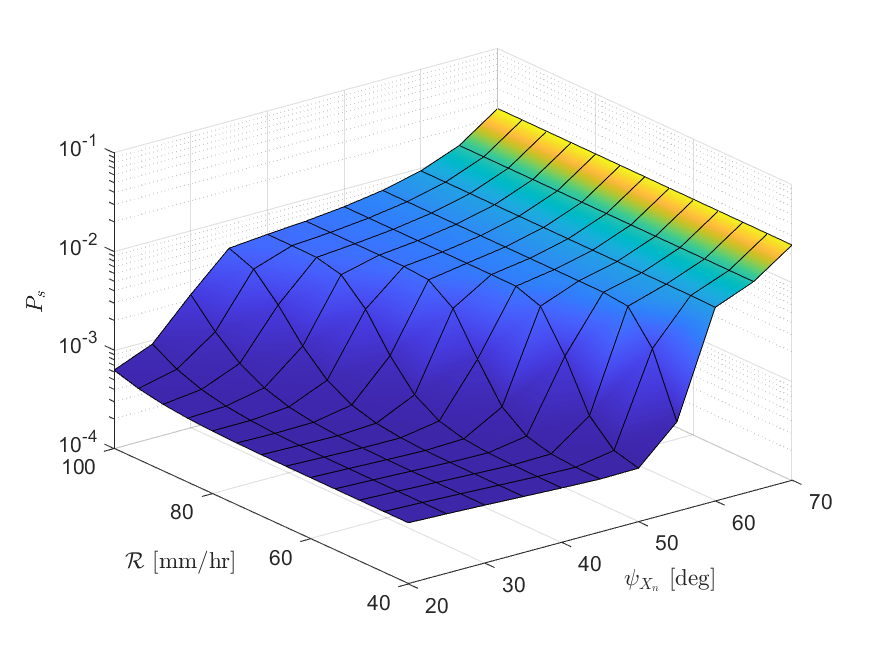}
         \caption{$h_{X_n}=19$ km.}
     \end{subfigure}
\begin{subfigure}[b]{0.5\textwidth}
     \label{case22}
         \centering
         \includegraphics[scale=0.53]{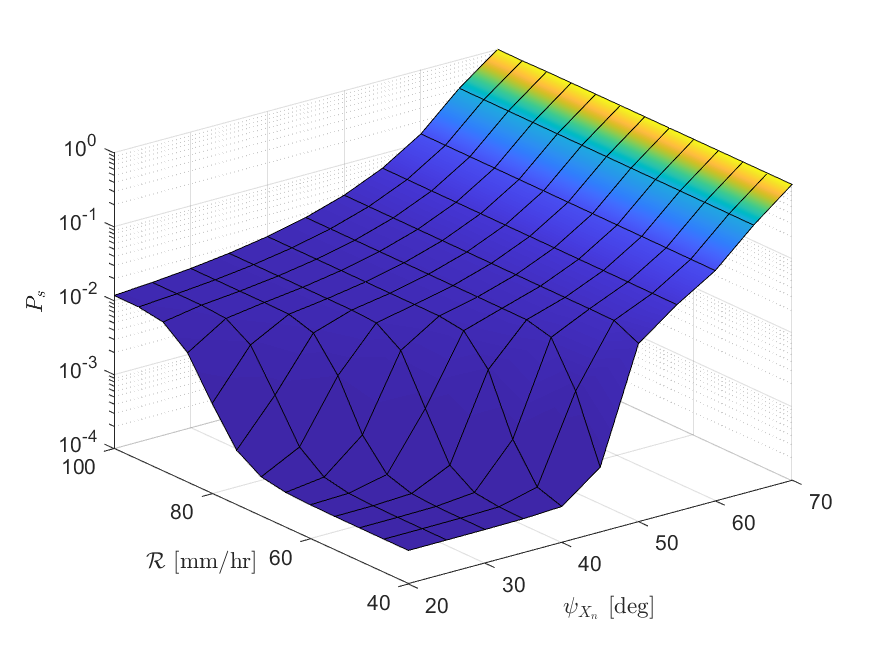}
         \caption{$h_{X_n}=40$ km.}
     \end{subfigure}
\caption{SOP\ of the proposed scheme vs. the rain rate $(\mathcal{R})$ and the zenith angle $\psi_{X_n}$.}
\label{figrain}
\end{figure}

In Fig. \ref{figrain}, the proposed scheme's performance is shown as a
function of different levels of precipitation rates and zenith angles. It is
worth mentioning that the curves were obtained by setting\ $R_{s}=5$ bps/Hz
and $\rho _{H_{n}}=3/4$, while the SOP\ performance is shown versus $%
\Upsilon _{H}$, with $\Upsilon _{U}$ [dB]=$\Upsilon _{H}$ [dB] $+30$, $%
\Upsilon _{E^{(1)}}=130$ dB, and $\Upsilon _{E^{(2)}}=160$ dB. One can note
the secrecy degradation of the considered scheme with the increase in the
rain rate and zenith angle, where a SOP\ level of $2.4\%$\ can be reached at
a precipitation rate $\mathcal{R=}100$ mm/hr (i.e., heavy rain) with a
zenith angle $\psi _{SX_{n}}=70^{\circ }$ at $h_{X_{n}}=19$ km, while a SOP\
of unit (i.e., worst secrecy scenario is reached at $h_{X_{n}}=40$ km at the
same aforementioned rain rate and zenith angle levels. Thus, the average
SNRs\ of the FSO, THz, and RF\ links are attenuated, yielding a decrease in
the respective link's SCs and an increase in their SOPs, degrading the total
SOP. It is worth mentioning that the rain attenuation effect is more
significant in zenith angle values between $20^{\circ }$ and $60^{\circ }$.

\begin{figure}[tbp]
\centering
\begin{subfigure}[b]{0.5\textwidth}
     \label{case11}
         \centering
         \includegraphics[scale=0.53]{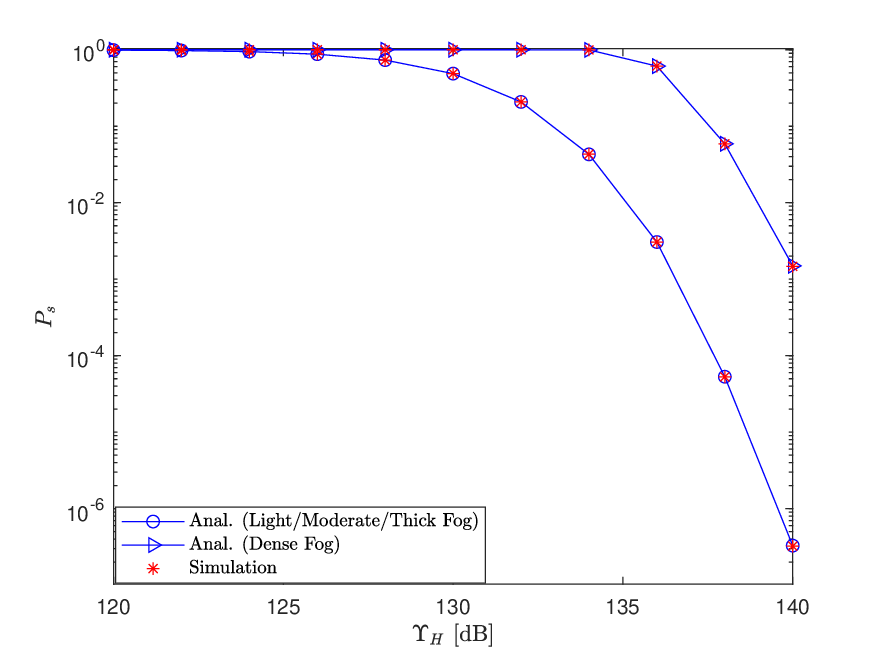}
         \caption{$\psi_{X_n}=30^{\circ}$.}
     \end{subfigure}
\begin{subfigure}[b]{0.5\textwidth}
     \label{case22}
         \centering
         \includegraphics[scale=0.53]{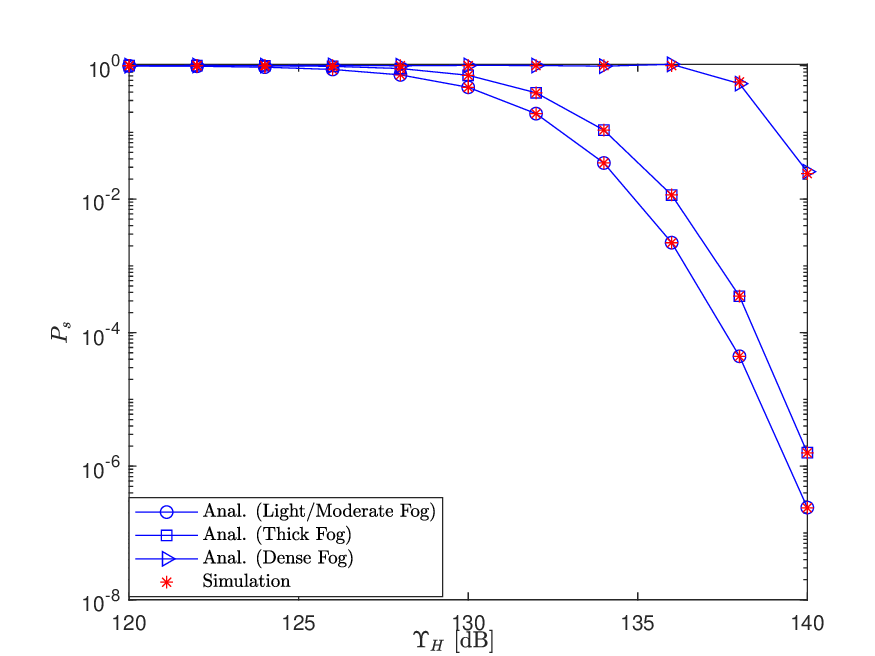}
         \caption{$\psi_{X_n}=50^{\circ}$.}
     \end{subfigure}
\caption{SOP\ of the proposed scheme vs. $\Upsilon_H$ for different fog regimes.}
\label{figfog}
\end{figure}

Fig. \ref{figfog} depicts the analyzed scheme's secrecy performance for
different levels of fog. The figure's curves were obtained by setting\ $%
R_{s}=5$ bps/Hz and $\rho _{H_{n}}=3/4$. Results show that the system's secrecy exhibits a significant degradation at a dense fog regime for a zenith angle $\psi_{X_n}=50^{\circ}$, where the secrecy loss can reach around 4 and 5 dB with respect to the thick, light, and moderate types of fog. It is worth noting also that at a lower zenith angle value, i.e., $\psi_{X_n}=30^{\circ}$, the scheme's secrecy is equal for light, moderate, and thick fog types, while it shows a secrecy loss of 3 dB with respect to the dense fog type. This is due to the fact that at lower $\psi_{X_n}$, the free-space path loss is less significant, and the transmit SNR can compensate for the fog loss and the free-space path loss at both the legitimate and illegitimate HAPs.

\section{Conclusion}

 In this paper, a secrecy-enhancing scheme for a HAP-aided hybrid
terrestrial-aerial transmission network was proposed. In particular, the
proposed scheme relied on the adoption of a THz feeder link as a backup for
the FSO\ one, whenever the latter one fails in achieving a target SC level.
In addition, the proposed scheme utilizes multi-HAP diversity to combat
eavesdropping through the presence of multiple relaying HAPs. A closed-form
expression for the overall system's\ SOP was derived encompassing the key
parameters of the considered scheme. Obtained results manifested the secrecy
enhancement of the proposed scheme with the increase of the FSO\ and THz
beam spot size at the receiver when transmitting in light or moderate RF\
signals shadowing, or the decrease of HAPs' zenith angles with respect to the
GS at higher altitudes. Furthermore, it has been shown that the proposed
scheme exhibits higher secrecy compared to the benchmark single-HAP\ and
FSO-based feeder link scheme, where both (i)\ HAPs diversity and (ii)\ the
presence of the backup THz link significantly contribute to enhancing the
security level of the network.\ Precisely, results manifested a SOP\
decrease by $N$ orders of magnitude at the high SNR level when raising the
number of HAPs from a unit to $N$. Furthermore, a $5$-dB secrecy gain is
manifested with respect to the benchmark FSO-based one at high SNR\ levels.

\section*{Appendix A:\ Proof of Proposition 1}

The proposed scheme is based on switching between the THz and FSO links per
their respective SCs, where the FSO\ link is primarily adopted whenever its
SC\ exceeds a communication threshold rate $R_{s}$. Otherwise, the THz link
is activated if its SC surpasses $R_{s}$. Therefore, an outage event takes
place when both links' SCs are less than $R_{s}$. Thus, we can formulate the
SOP of the first hop link for the pair of legitimate/illegitimate HAPs,
i.e., $H_{n}/E_{n},$ as
\begin{equation}
P_{s}^{(1,n)}=\Pr \left[ C_{s,1,\text{FSO}}^{(n)}\leq R_{s},C_{s,1,\text{THz}%
}^{(n)}\leq R_{s}\right]  \label{sop1init}
\end{equation}%
As both the FSO\ and THz links are independent, the expression in (\ref%
{sop1init}) can be reformulated as%
\begin{equation}
P_{s}^{(1,n)}=\underset{\triangleq P_{s,\text{FSO}}^{(1,n)}}{\underbrace{\Pr %
\left[ C_{s,1,\text{FSO}}^{(n)}\leq R\right] }}\underset{\triangleq P_{s,%
\text{THz}}^{(1,n)}}{\underbrace{\Pr \left[ C_{s,1,\text{THz}}^{(n)}\leq
R_{s}\right] }}  \label{sop1final}
\end{equation}

On the other hand, when DF\ relaying is implemented, it is established that
the e2e SC of DF-based dual-hop wireless networks is the minimum of the two
hops' SCs. Henceforth, an outage in at least one of the two hops results in
an e2e outage event, Thus, for a given pair $H_{n}/E_{n}$, the e2e SOP\ is
expressed a%
\begin{align}
P_{s}^{(n)} &=P_{s}^{(1,n)}\left( 1-P_{s,\text{Ka}}^{(2,n)}\right) +\left(
1-P_{s}^{(1,n)}\right) P_{s,\text{Ka}}^{(2,n)}  +P_{s}^{(1,n)}P_{s,\text{Ka}}^{(2,n)}  \notag \\
&=P_{s}^{(1,n)}+P_{s,\text{Ka}}^{(2,n)}-P_{s}^{(1,n)}P_{s,\text{Ka}%
}^{(2,n)},  \label{sopntot}
\end{align}%
where%
\begin{equation}
P_{s,\text{Ka}}^{(2,n)}=\Pr \left[ C_{s,2,\text{Ka}}^{(n)}\leq R_{s}\right]
\label{sop2}
\end{equation}%
is the second hop's SC. As a result, the e2e\ SOP\ of the dual-hop link
through the $n$th HAP is expressed from (\ref{sop1final})-(\ref{sop2})\ as%
\begin{equation}
P_{s}^{(n)}=P_{s,\text{FSO}}^{(1,n)}P_{s,\text{THz}}^{(1,n)}+P_{s,\text{Ka}%
}^{(2,n)}-P_{s,\text{FSO}}^{(1,n)}P_{s,\text{THz}}^{(1,n)}P_{s,\text{Ka}%
}^{(2,n)},  \label{sopntotfinal}
\end{equation}%
Finally, as the proposed scheme picks the HAP exhibiting the greatest SC\
(i.e., lowest SOP), it can be inferred that a secrecy outage event occurs
only if all the $N$ links (through the $N$ HAPs)\ are in outarge, that is%
\begin{align}
P_{s}& =\prod\limits_{n=1}^{N}P_{s}^{(n)}  \notag \\
& =\prod\limits_{n=1}^{N}\left[ P_{s,\text{FSO}}^{(1,n)}P_{s,\text{THz}%
}^{(1,n)}+P_{s,\text{Ka}}^{(2,n)}-P_{s,\text{FSO}}^{(1,n)}P_{s,\text{THz}%
}^{(1,n)}P_{s,\text{Ka}}^{(2,n)}\right]
\end{align}%
which concludes the proposition's proof.

\section*{Appendix B:\ Proof of Proposition 2}

\subsection{\protect FSO\ Link}

 The SOP of the $n$th FSO\ link (i.e., $S$-$H_{n}$) can be expressed
by inserting the CDF\ in (\ref{cdffso}) with $X=H_{n}$ and the PDF\ in (\ref%
{pdffso}) with $X=E^{(1)}$ into (\ref{sopdefintegral}) as%
\begin{align}
F_{\gamma _{1,X_{n}}^{\text{(FSO)}}}(z)& =\frac{r^{\alpha _{X_{n}}+\beta
_{X,n}-2}\xi _{X_{n},\text{FSO}}^{2}}{\left( 2\pi \right) ^{r-1}\Gamma
\left( \alpha _{X_{n}}\right) \Gamma \left( \beta _{X,n}\right) }  \notag \\
& \times G_{r+1,3r+1}^{3r,1}\left( \frac{\left( \mathcal{G}_{X_{n}}\right)
^{r}z}{\overline{\gamma }_{X_{n}}^{\text{(FSO)}}}\left\vert
\begin{array}{c}
1;\varepsilon _{1}^{(X_{n})} \\
\varepsilon _{2}^{(X_{n})};0%
\end{array}%
\right. \right) ,
\end{align}%
\begin{subequations}
\begin{align}
& P_{s,\text{FSO}}^{(1,n)}\overset{(a)}{=}\frac{\xi _{E_{n},\text{FSO}}^{2}}{%
r\Gamma \left( \alpha _{E_{n}}\right) \Gamma \left( \beta _{E_{n}}\right) }%
\frac{r^{\alpha _{H_{n}}+\beta _{H_{n}}-2}\xi _{H_{n},\text{FSO}}^{2}}{%
\left( 2\pi \right) ^{r-1}\Gamma \left( \alpha _{H_{n}}\right) \Gamma \left(
\beta _{H_{n}}\right) }  \notag \\
& \times \int_{0}^{\infty }G_{1,3}^{3,0}\left( \mathcal{F}_{E_{n},1}\left(
\frac{z}{\overline{\gamma }_{E_{n}}^{\text{(FSO)}}}\right) ^{\frac{1}{r}%
}\left\vert
\begin{array}{c}
-;\xi _{E_{n},\text{FSO}}^{2}+1 \\
\xi _{E_{n},\text{FSO}}^{2},\alpha _{E_{n}},\beta _{E_{n}};-%
\end{array}%
\right. \right)  \notag \\
& \times G_{r+1,3r+1}^{3r,1}\left( \frac{\mathcal{F}_{H_{n}^{(1)},r}z}{%
\overline{\gamma }_{H_{n}}^{\text{(FSO)}}}\left\vert
\begin{array}{c}
1;\varepsilon _{1}^{(H_{n})} \\
\varepsilon _{2}^{(H_{n})};0%
\end{array}%
\right. \right) \\
& \overset{(b)}{=}\frac{\xi _{E_{n},\text{FSO}}^{2}\xi _{H_{n},\text{FSO}%
}^{2}\left( 2\pi \right) ^{1-r}r^{\alpha _{H_{n}}+\beta _{H_{n}}-3}}{\Gamma
\left( \alpha _{E_{n}}\right) \Gamma \left( \alpha _{H_{n}}\right) \Gamma
\left( \beta _{E_{n}}\right) \Gamma \left( \beta _{H_{n}}\right) }\frac{1}{%
2\pi j}  \notag \\
& \int_{C_{s}}\left[ \mathcal{F}_{X_{n},r}\frac{2^{R}}{\overline{\gamma }%
_{H_{n}}^{\text{(FSO)}}}\right] ^{-s}\mathcal{T}_{H_{n}}(r,s)\int_{0}^{%
\infty }\frac{z^{-1}}{\left( z+1-\frac{1}{2^{R}}\right) ^{s}}  \notag \\
& \times G_{1,3}^{3,0}\left( \mathcal{F}_{E_{n},1}\left( \frac{z}{\overline{%
\gamma }_{E_{n}}^{\text{(FSO)}}}\right) ^{\frac{1}{r}}\left\vert
\begin{array}{c}
-;\xi _{E_{n},\text{FSO}}^{2}+1 \\
\xi _{E_{n},\text{FSO}}^{2},\alpha _{E_{n}},\beta _{E_{n}};-%
\end{array}%
\right. \right) ds  \label{sopfsosteps}
\end{align}%
with
\end{subequations}
\begin{align}
\mathcal{T}_{H_{n}}(r,s) &=\frac{
\prod\limits_{i=0}^{r-1}\Gamma \left( \frac{\xi _{H_{n},\text{FSO}}^{2}+i}{r}%
+s\right) \Gamma \left( \frac{\alpha _{H_{n}}+i}{r}+s\right) \Gamma \left( \frac{\beta _{H_{n}}+i}{r}+s\right)}{\left(
-s\right) \prod\limits_{i=1}^{r}\Gamma \left( \frac{\xi _{H_{n},\text{FSO}%
}^{2}+i}{r}\right) }  
\end{align}%
where Step (b)\ in (\ref{sopfsosteps}) stems from Step (a) through the
Mellin-Barnes representation of the Meijer's $G$-function in \cite[Eq.
(07.34.02.0001.01)]{wolfram}. Thus, relying on the identity \cite[Eq.
(07.34.21.0086.01)]{wolfram} and some algebraic manipulations, it yields
\begin{align}
P_{s,\text{FSO}}^{(1,n)}& =\frac{\xi _{E_{n},\text{FSO}}^{2}\xi _{H_{n},%
\text{FSO}}^{2}r^{\alpha _{E_{n}}+\beta _{E_{n}}+\alpha
_{H_{n}}+\beta _{H_{n}}-3}}{r \left( 2\pi \right) ^{2\left( r-1\right) }\Gamma \left( \alpha _{E_{n}}\right) \Gamma \left( \beta
_{E_{n}}\right) \Gamma
\left( \alpha _{H_{n}}\right) \Gamma \left( \beta _{H_{n}}\right) } \left( \frac{1}{2\pi j}\right) ^{2} \notag
\\
& \times \int_{C_{s}}\int_{C_{v}}\mathcal{%
T}_{H_{n}}(r,s)\Gamma \left( 1-v\right) \mathcal{T}_{E_{n}}(r,v)
 Q_{H_{n}}^{-s}Q_{E_{n}}^{-v}dsdv .\label{sopfsolast2}
\end{align}%
It is worthwile that the double complex integral above requires the
following conditions on the complex contours of integral, i.e., $C_{s}=\left]
c_{s}-j\infty ,c_{s}+j\infty \right[ ,$ $C_{v}=\left] c_{v}-j\infty
,c_{v}+j\infty \right[ $, as%
\begin{equation}
c_{s}\in \left] -\frac{\min \left( \xi _{H_{n},\text{FSO}}^{2},\alpha
_{H_{n}},\beta _{H_{n}}\right) }{2},0\right[ \Longleftrightarrow c_{s}<0,
\end{equation}%
which implies that the poles of $\Gamma \left( -v\right) $ and $\Gamma
\left( s+v\right) $ will overlap as $c_{v}>0$ to guarantee a non-overlapping
of the poles of the second integral. Thus, residues on poles of $\Gamma
\left( -v\right) $ from the first pole $v=0$ up to $c_{v}$ must be
subtracted from the double integral term in (\ref{sopfsolast2}). Therefore,
relying on the aforementioned subtraction\ and the bivariate Meijer's $G$%
-function representation in \cite{yakub}, it yields (\ref{sopfsofinal}).

\subsection{THz Link}

By incorporating the CDF\ in (\ref{cdfthz}) with $X=H_{n}$ and the PDF\ in (%
\ref{pdfthz}) with $X=E_{n}$ into (\ref{sopdefintegral}), the SOP\ of the
first hop's $n$th link ($S$-$H_{n}$)\ link when operating with the THz band
can be expressed as%
\begin{equation}
P_{s,\text{THz}}^{(1,n)}=\int_{0}^{\overline{\gamma }_{1,E_{n}}^{\text{%
(THz,max)}}}F_{\gamma _{SH_{n}}^{\text{(THz)}}}\left( 2^{R_{s}}\left(
z+1\right) -1\right) f_{\gamma _{SE_{n}}^{\text{(THz)}}}\left( z\right) dz.
\label{sopthz1}
\end{equation}%
where $\gamma _{1,X_{n}}^{\text{(THz,max)}}$ and $\Psi _{1,H_{n}}$ are
defined in the proposition. It can be observed that the integral in the last
equation can exhibit three cases, per the intervals defining the legitimate
link's CDF as:

\subsubsection{Case 1:\ $\protect\gamma _{1,E_{n}}^{\text{(THz,max)}}>\Psi
_{1,H_{n}}>0$}

In such a scenario, the integral in (\ref{sopthz1}) can be expressed as the
sum of two terms, as the legitimate link's CDF term in (\ref{sopthz1})
equals unit for $z>\Psi _{1,H_{n}}$, as can be deduced from (\ref{cdfthz}).
Thus, it yields
\begin{subequations}
\begin{align}
P_{s,\text{THz}}^{(1,n)} &\overset{(a)}{=}\frac{\left( 2^{R_{s}}-1\right) ^{%
\frac{\xi _{H_{n},\text{THz}}^{2}}{2}}\xi _{E_{n},\text{THz}}^{2}\left(
\overline{\gamma }_{1,H_{n}}^{\text{(THz)}}\right) ^{-\frac{\xi _{H_{n},%
\text{THz}}^{2}}{2}}}{2\left( A_{0}^{\text{(THz)}}\right) ^{\xi _{H_{n},%
\text{THz}}^{2}+\xi _{E,n,\text{THz}}^{2}}\left( \overline{\gamma }%
_{1,E_{n}}^{\text{(THz)}}\right) ^{\frac{\xi _{E_{n},n,\text{THz}}^{2}}{2}}}
\notag \\
&\times \int_{0}^{\Psi _{1,H_{n}}}\frac{z^{\frac{\xi _{E_{n},\text{THz}}^{2}%
}{2}-1}}{\left( \frac{2^{R_{s}}}{2^{R_{s}}-1}z+1\right) ^{-\frac{\xi _{H_{n},%
\text{THz}}^{2}}{2}}}dz  \notag \\
&+\frac{\xi _{E_{n},n,\text{THz}}^{2}}{2\left( A_{0}^{\text{(THz)}}\right)
^{\xi _{E_{n},\text{THz}}^{2}}\overline{\gamma }_{1,E_{n}}^{\text{(THz)}}}%
\int_{\Psi _{1,H_{n}}}^{\gamma _{1,E_{n}}^{\text{(THz,max)}}}\left( \frac{z}{%
\overline{\gamma }_{1,E_{n}}^{\text{(THz)}}}\right) ^{\frac{\xi _{E_{n},n,%
\text{THz}}^{2}}{2}-1} \\
&\overset{(b)}{=}\frac{\left( 2^{R_{s}}-1\right) ^{\frac{\xi _{H_{n},\text{%
THz}}^{2}}{2}}\xi _{E,n,\text{THz}}^{2}\left( \overline{\gamma }_{1,H_{n}}^{%
\text{(THz)}}\right) ^{-\frac{\xi _{H,n,\text{THz}}^{2}}{2}}}{2\left( A_{0}^{%
\text{(THz)}}\right) ^{\xi _{H,n,\text{THz}}^{2}+\xi _{E,n,\text{THz}%
}^{2}}\left( \overline{\gamma }_{1,E_{n}}^{\text{(THz)}}\right) ^{\frac{\xi
_{E_{n},n,\text{THz}}^{2}}{2}}}  \notag \\
&\times \text{ }_{2}F_{1}\left( -\frac{\xi _{H_{n},\text{THz}}^{2}}{2},%
\frac{\xi _{E_{n},\text{THz}}^{2}}{2};\frac{\xi _{E_{n},\text{THz}}^{2}}{2}%
+1;-\frac{2^{R_{s}}\Psi _{1,H_{n}}}{2^{R_{s}}-1}\right)  \notag \\
&+F_{\gamma _{SE_{n}}^{\text{(THz)}}}\left( \gamma _{1,E_{n}}^{\text{%
(THz,max)}}\right) -F_{\gamma _{SE_{n}}^{\text{(THz)}}}\left( \Psi
_{1,H_{n}}\right)
\end{align}%
where Step (b)\ is reached by utilizing \cite[Eq. (3.194.1)]{integrals} for
computing the first term, while the second term of Step (a)\ yields the
difference of CDF\ values at the two specified integration bounds, i.e., $%
\Psi _{1,H_{n}}$ and $\gamma _{1,E_{n}}^{\text{(THz,max)}}$.

\subsubsection{Case 2:\ $\protect\gamma _{1,E_{n}}^{\text{(THz,max)}}<\Psi
_{1,H_{n}}$}

In this case, the integral of (\ref{sopthz1}) is truncated to $\gamma
_{1,E_{n}}^{\text{(THz,max)}}$ as the PDF\ vanishes for values exceeding the
aforementioned bound. Therefore, this yields, similarly to Case 1, the
\end{subequations}
\begin{align}
P_{s,\text{THz}}^{(1,n)}& =\frac{\left( 2^{R_{s}}-1\right) ^{\frac{\xi
_{H_{n},\text{THz}}^{2}}{2}}\xi _{E,n,\text{THz}}^{2}\left( \overline{\gamma
}_{1,H_{n}}^{\text{(THz)}}\right) ^{-\frac{\xi _{H,n,\text{THz}}^{2}}{2}}}{%
2\left( A_{0}^{\text{(THz)}}\right) ^{\xi _{H,n,\text{THz}}^{2}+\xi _{E,n,%
\text{THz}}^{2}}\left( \overline{\gamma }_{1,E_{n}}^{\text{(THz)}}\right) ^{%
\frac{\xi _{E_{n},n,\text{THz}}^{2}}{2}}}  \notag \\
& \times \text{ }_{2}F_{1}\left( -\frac{\xi _{H_{n},\text{THz}}^{2}}{2},%
\frac{\xi _{E_{n},\text{THz}}^{2}}{2};\frac{\xi _{E_{n},\text{THz}}^{2}}{2}%
+1;-\frac{2^{R_{s}}\gamma _{1,E_{n}}^{\text{(THz,max)}}}{2^{R_{s}}-1}\right)
\end{align}

\subsubsection{\ Case 3:\ $\Psi _{1,H_{n}}<0$}

This third case is manifested when the CDF's argument in (\ref{sopthz1}) is
always above the integral's upper bound for all integration values , i.e., $%
2^{R_{s}}\left( z+1\right) -1>\gamma _{1,E_{n}}^{\text{(THz,max)}}$ $\forall
z>0$. Thus, $F_{\gamma _{SH_{n}}^{\text{(THz)}}}\left( 2^{R_{s}}\left(
z+1\right) -1\right) =1$ for $z\in \left[ 0,\gamma _{1,E_{n}}^{\text{%
(THz,max)}}\right] $ and the SOP\ equals unit as it is the integral of the
PDF $f_{\gamma _{SE_{n}}^{\text{(THz)}}}\left( z\right) $\ over the whole
range of $\gamma _{SE_{n}}^{\text{(THz)}}$.

Therefore, by grouping the three above-detailed cases together, this yields (%
\ref{sopthzfinal}).

\subsection{RF\ link}

Similarly to the FSO\ and THz links, the SOP\ of the $H_{n}$-$U$ RF\ link
yields from incorporating (\ref{cdfka}) with $Z=U$ and (\ref{pdfka}) with $%
Z=E^{(2)}$ into (\ref{sopdefintegral}) as follows
\begin{subequations}
\begin{align}
 P_{s,\text{Ka}}^{(2,n)}&\overset{(a)}{=}\frac{\phi _{U}\phi _{E^{(2)}}}{\overline{%
\gamma }_{2,E^{(2)}}}\sum\limits_{n=0}^{m_{s}^{(E^{(2)})}-1}\frac{\binom{%
m_{s}^{(E^{(2)})}-1}{n}\left( \frac{\mu _{E^{(2)}}}{\overline{\gamma }%
_{2,E^{(2)}}}\right) ^{n}}{n!}  \sum\limits_{p=0}^{m_{s}^{(U)}-1}%
\frac{\mu _{U}^{p}}{v_{U}^{p+1}} \notag \\
& \times \frac{\binom{m_{s}^{(U)}-1}{p}}{p!}   \int_{0}^{\infty } \frac{z^{n}\gamma _{\text{inc}}\left( p+1,\frac{\mu _{U}\left(
2^{R_{s}}\left( 1+z\right) -1\right) }{\overline{\gamma }_{2,U}}\right)}{\exp \left( \frac{%
v_{E^{(2)}}z}{\overline{\gamma }_{2,E^{(2)}}}\right)} dz,
\\
& \overset{(b)}{=}\frac{\phi _{U} \phi _{E^{(2)}}\exp \left( -\frac{v_{E^{(2)}}\left[
\frac{1}{2^{R_{s}}}-1\right] }{\overline{\gamma }_{2,E^{(2)}}}\right) }{%
2^{R_{s}\left( n+1\right) }\overline{\gamma }_{2,E^{(2)}}}%
\sum\limits_{n=0}^{m_{s}^{(E^{(2)})}-1}\frac{\binom{m_{s}^{(E^{(2)})}-1}{n}}{%
n!}  \notag \\
& \times \left( \frac{\mu _{E^{(2)}}}{\overline{\gamma }_{2,E^{(2)}}}\right)
^{n}\sum\limits_{p=0}^{m_{s}^{(U)}-1}\frac{\binom{m_{s}^{(U)}-1}{p}%
\mu _{U}^{p}}{v_{U}^{p+1}} \times \sum\limits_{q=0}^{n}\frac{\binom{n}{q}}{\left( 1-2^{R_{s}}\right)
^{q-n}} \notag \\
& \int_{2^{R_{s}}-1}^{\infty }\frac{\exp \left( -\frac{v_{E^{(2)}}t}{\overline{%
\gamma }_{2,E^{(2)}}2^{R_{s}}}\right) }{t^{-q}} \left[ 1-e^{-\frac{\mu _{U}t}{\overline{\gamma }_{2,U}}%
}\sum\limits_{k=0}^{p}\frac{\left( \frac{\mu _{U}t}{\overline{\gamma }_{2,U}}%
\right) ^{k}}{k!}\right] dt,
\end{align}  
\end{subequations}
where Step (b) can be achieved from Step (a)\ by virtue of the change of
variable $t=2^{R_{s}}\left( 1+z\right) -1$ the binomial expansion for the
term $\left( t+1-2^{R}\right) ^{n}$, and the identity \cite[Eq. (8.352.1)]%
{integrals}. Finally, using \cite[Eq. (3.381.3)]{integrals}, (\ref%
{soprffinal}) is reached.

\bibliographystyle{IEEEtran}
\bibliography{refs}

\end{document}